\documentclass[%
reprint,
aps,
prb,
]{revtex4-1}

\usepackage{graphicx}
\usepackage{dcolumn}
\usepackage{bm}

\usepackage{mathrsfs}
\usepackage{multirow}
\usepackage{booktabs}

\newcommand{\V}[1]{\bm{#1} } 

\newcommand{\Ave}[1]{\left\langle {#1} \right\rangle} 
 
\newcommand{\sgn}[1]{{\rm sgn}\left({#1} \right)} 
\newcommand{\lsim}{\ $\raisebox{-.7ex}{$\stackrel{\textstyle <}{\sim}$}$\,\ }
\newcommand{\gsim}{\ $\raisebox{-.7ex}{$\stackrel{\textstyle >}{\sim}$}$\,\ }

\newcommand{\lb}{\left(}
\newcommand{\rb}{\right)}

\newcommand{\lsb}{ \left[ }
\newcommand{\rsb}{ \right] }

\newcommand{\T}[1]{\tilde{#1}}
\newcommand{\Req}[1]{Eq.\ (\ref{eq:#1})}

\newcommand{\Reqs}[2]{Eqs.\ (\ref{eq:#1},\ref{eq:#2})}

\newcommand{\Rfig}[1]{Fig.\ \ref{fig:#1}}
\newcommand{\Rfigss}[2]{Figs.\ \ref{fig:#1}-\ref{fig:#2}}
\newcommand{\Lfig}[1]{\label{fig:#1}}
\newcommand{\Leq}[1]{\label{eq:#1}}
\newcommand{\Rsec}[1]{Sec.\ \ref{sec:#1}}
\newcommand{\Lsec}[1]{\label{sec:#1}}
\newcommand{\be}{\begin{eqnarray}}
\newcommand{\ee}{\end{eqnarray}}
\newcommand{\ba}{\begin{array}}
\newcommand{\ea}{\end{array}}

\newcommand{\subbe}{\begin{subequations}}
\newcommand{\subee}{\end{subequations}}

\begin{document}

\preprint{APS/123-QED}

\title{Monte Carlo simulations of the three-dimensional {\it XY\/} spin glass focusing on the chiral and the spin order}

\author{Tomoyuki Obuchi}
\author{Hikaru Kawamura}%
\affiliation{%
Department of Earth and Space Science, Graduate School of Science, Osaka University, Toyonaka, Osaka, 560-0043, Japan,
}%




\date{\today}

\begin{abstract}
 The ordering of the three-dimensional isotropic {\it XY} spin glass with the nearest-neighbor random Gaussian coupling is studied by extensive Monte Carlo simulations. To investigate the ordering of the spin and the chirality, we compute several independent physical quantities including the glass  order parameter, the Binder parameter, the correlation-length ratio, the overlap distribution and the non-self-averageness parameter, {\it etc\/}, for both the spin-glass (SG) and the chiral-glass (CG) degrees of freedom. Evidence of the spin-chirality decoupling, {\it i.e.\/}, the CG and the SG order occurring at two separated temperatures, $0<T_{SG}<T_{CG}$, is obtained from the glass order parameter, which is fully corroborated by the Binder parameter. By contrast, the CG correlation-length ratio yields a rather pathological and inconsistent result in the range of sizes we studied, which may originate from the finite-size effect associated with a significant short-length drop-off of the spatial CG correlations. Finite-size-scaling analysis yields the CG exponents $\nu_{CG}=1.36^{+0.15}_{-0.37}$ and $\eta_{CG}=0.26^{+0.29}_{-0.26}$, and the SG exponents $\nu_{SG}=1.22^{+0.26}_{-0.06}$ and $\eta_{SG}=-0.54^{+0.24}_{-0.52}$. The obtained exponents are close to those of the Heisenberg SG, but are largely different from those of the Ising SG. The chiral overlap distribution and the chiral Binder parameter exhibit the feature of a continuous one-step replica-symmetry breaking (1RSB), consistently with the previous reports. Such a 1RSB feature is again in common with that of the Heisenberg SG, but is different from the Ising one, which may be the cause of the difference in the CG critical properties from the Ising SG ones despite of a common $Z_2$ symmetry.


\end{abstract}

\maketitle


\section{\label{sec:intro}INTRODUCTION}

In spite of long history of research, spin glass (SG) is still a  hot topic in statistical physics~\cite{review}. It is a typical system possessing both strong frustration and randomness, leading to several extraordinary behaviors such as the slow dynamics and the rejuvenation-memory effect. In a theoretical treatment of SG, Edwards and Anderson (EA) proposed as early as in 1972 a simple model~\cite{Edwards:75}, the so-called EA model, in which the spins are put on each site of a regular lattice and interact via the random coupling taking both positive and negative signs. The infinite-range or the mean-field version of the EA model, first presented by Sherrington and Kirkpatrick (SK)~\cite{Sherrington:75}, was solved by Parisi revealing an intriguing concept of the replica symmetry breaking (RSB)~\cite{Parisi:79}. For both cases of the Ising and the Heisenberg SK models, the relevant RSB turned out to be of hierarchical nature. In spite of such success of the mean-field theory, understanding the nature of the ordering of the finite-range EA model in three dimensions (3D) still remains incomplete. Numerical simulations have been the main tool in attacking the issue. Although the existence of a finite-temperature SG transition was established in the 3D Ising EA model~\cite{Ogielski:85,Bhatt:85,Ballesteros:00,Katzgraber:06,Jorg:06,Campbell:06,Hasenbusch:08}, earlier numerical simulations on the 3D {\it XY\/} and the Heisenberg EA models  suggested the absence of a finite-temperature transition~\cite{Banavar:84,McMillan:85,Olive:85,Morris:86,Jain:86,Matsubara:91,Yoshino:91}, in apparent contrast to experiments~\cite{Simpson:79,Courenay:86,Bouchiat:86,Levy:86,Coles:88,Fert:88,Taniguchi:88,Campbell:10}.

 Some time ago,  in discussing the ordering of frustrated vector spin systems, Villain proposed a possible significance of the ``chirality'' degree of freedom, an Ising-like scalar quantity which represents the handedness of the noncollinear spin structure. Villain made a conjecture that the 3D {\it XY\/} SG might exhibit a finite-temperature SG ordering by noting the Ising nature of the chirality and by invoking the occurrence of a finite-temperature SG transition in the 3D Ising SG~\cite{Villain:77}.

 Numerical simulations on the vector SG, {\it i.e.\/}, the two-component {\it XY\/} SG or the three-component Heisenberg SG, investigating both the spin and the chirality degrees of freedom, have been performed ever since 1985~\cite{Kawamura:85}. A crucially important concept which emerged from these studies is the possible ``spin-chirality decoupling'' phenomenon~\cite{Kawamura:92,Kawamura:07,Kawamura:10}. In 3D, it means that the chirality orders at a temperature higher than the spin, with an intermediate ``chiral-glass'' (CG) phase where only the chirality exhibits a glassy long-range order while the standard SG order still remains short-ranged. In dimensions lower than three, where both the spin and the chirality are believed to order only at $T=0$, the spin-chirality decoupling means that the spin and the chiral correlation-length exponents are mutually different, {\it i.e.\/}, the existence of two different diverging length scales at the $T=0$ transition~\cite{Kawamura:87,Batrouni:88,Horiguchi:90,Kawamura:91,Ray:92,Ney-Nifle:93,Ney-Nifle:95,Bokil:96,Wengel:97,Granato:98,Kosterlitz:99,Kawamura:03-2,Uda:05,Weigel:08}.  In terms of a symmetry, in the CG phase the $Z_2$ spin-reflection symmetry is spontaneously broken with keeping the $SO(3)$ or $SO(2)$ spin-rotation symmetry unbroken.
 
 In case of the 3D {\it XY\/} SG, which is a target of the present study, the spin-chirality decoupling was first examined by the numerical domain-wall renormalization-group calculation and also by a Monte Carlo (MC) simulation~\cite{Kawamura:91,Kawamura:95-1,Kawamura:01-1}. Further interesting features revealed by these numerical analyses might be the possible one-step RSB (1RSB)-like feature of the CG ordered state, and the non-Ising character of the CG criticality~\cite{Kawamura:01-1}. 

 Similar spin-chirality decoupling phenomena, the 1RSB-like nature of the CG ordered state and the non-Ising character of the CG criticality were also observed in the 3D Heisenberg SG, a reference model of many of realistic SG materials including canonical SG, in spite of the difference in the nature of the chiralities relevant to the {\it XY\/} and the Heisenberg spins. For the former, the chirality is quadratic in spins and time-reversal even, while, in the latter, it is cubic in spins and time-reversal odd. Indeed, on the basis of the spin-chirality decoupling picture of the 3D Heisenberg SG, the chirality scenario of the experimental SG order was advanced~\cite{Kawamura:92,Kawamura:07,Kawamura:10}. Recent large-scale simulations have revealed that the SG order actually takes place at a nonzero temperature~\cite{Kawamura:98,Kawamura:99,Kawamura:01-2,Imagawa:03,Kawamura:03-1,Imagawa:04,Hukushima:05,Viet:09,Viet:10,Sharma:11,Campos:06,Lee:07,Fernandez:09,Martin-Mayor:12}, in contrast to earlier beliefs. Besides, although some counter opinions are still present~\cite{Campos:06,Lee:07,Fernandez:09,Martin-Mayor:12}, several recent MC simulations give eloquent evidences that  the spin-chirality decoupling actually occurs, {\it i.e.\/}, $0<T_{SG}<T_{CG}$, in a range of dimensions including three~\cite{Kawamura:98,Kawamura:99,Kawamura:01-2,Imagawa:03,Kawamura:03-1,Imagawa:04,Hukushima:05,Viet:09,Viet:10,Sharma:11}. Several experimental facts were also successfully explained by the chirality scenario~\cite{Campbell:10,Kawamura:10}, which clearly emphasizes the physical importance of the spin-chirality decoupling in understanding the realistic SG systems.
 
On the other hand, the situation in the 3D {\it XY} SG seems less clear. Extensive calculations comparable in their scale to those of the 3D Heisenberg SG have been scarce, and the occurrence of the spin-chirality decoupling still remains controversial. Maucourt and Grempel suggested  on the basis of their $T=0$ domain-wall renormalization-group calculation for lattices $L\leq 8$ the occurrence of a nonzero $T_{SG}$ located below $T_{CG}$~\cite{Maucourt:98}. Mentioning some of the recent MC simulations on the model, Kawamura and Li simulated the $\pm J$ EA model by an equilibrium MC simulation up to the linear size $L=16$, suggesting the occurrence of the spin-chirality decoupling~\cite{Kawamura:01-1}, whereas Granato performed the dynamical Langevin simulations of the model for lattices of $L\leq12$, to conclude the occurrence of a single transition $T_{SG}=T_{CG}$~\cite{Granato:01,Granato:04}. Nakamura and collaborators performed the nonequilibrium relaxation analysis for lattices up to $L=55$ (this method enables one to treat relatively larger sizes but some drawbacks appear in its short-time observations), and suggested that $T_{SG}$ and $T_{CG}$ were identical or close even if they were to be different~\cite{Yamamoto:04,Nakamura:06}. Young and collaborators investigated the Gaussian EA model by equilibrium MC simulations for lattices up to $L\leq 24$, reporting no evidence of the spin-chirality decoupling~\cite{Lee:03,Pixley:08}.

This confusing situation motivates us to re-examine the ordering of the 3D {\it XY\/} EA model with the random Gaussian coupling by large-scale MC simulations by treating large sizes up to $L=40$, considerably larger than the sizes studied before by equilibrium simulations. A large number of samples of order $N_{s}\sim O(10^{3})$ are simulated to obtain reasonable statistics. Furthermore, we compute various independent quantities including the glass order parameter, the Binder parameter, the correlation-length ratio, the overlap-distribution function and  the non-self-averageness parameters both for the SG and the CG, in order to check consistency among various independent quantities. 

We note that the 3D {\it XY\/} EA model is a reference model for SG magnets with an easy-plane-type uniaxial magnetic anisotropy~\cite{Katsumata:88,Murayama:86,Mathieu:05,Ito:92,Yamaguchi:12,Kawamura:11}. Readers are referred to Ref. 74 for detailed discussion. The ordering properties of the model would also be helpful in understanding the peculiar ordering behaviors experimentally observed in these granular cuprate superconductors~\cite{Matsuura:95,Yamao:99,Papadopoulou:99,Gardchareon:03,Hagiwara:05,Deguchi:09,Kawamura:95-2,KawamuraLi:97}.

 Overall, the results of our large-scale simulations speak for the occurrence of the spin-chirality decoupling in the 3D {\it XY\/} SG. The estimated SG and CG transition temperatures are $T_s=0.275^{+ 0.013}_{-0.052}$ and $T_{CG}=0.313^{+0.013}_{-0.018}$, $T_{CG}$ being higher than $T_{SG}$ by about 10\%. In estimating the transition temperatures, we have found a reasonably good consistency among various independent quantities, with one exception of the CG correlation-length ratio, which behaves rather badly leading to a pathological estimate of $T_{CG}$. Thus, in deriving the above estimate of $T_{CG}$, we have not used the CG-correlation length data, in contrast to Ref. 67. To clarify the origin of the observed pathological behavior of the CG correlation length, we directly compute the spatial chiral correlation function to find that the standard formula of the finite-size correlation length may be inappropriate to describe the CG correlation length in the range of small sizes in which we make our simulations.

  The critical properties associated with the SG and the CG orderings are also examined. We obtain the CG exponents $\nu_{CG}=1.36^{+0.15}_{-0.37}$ and $\eta_{CG}=0.26^{+0.29}_{-0.26}$, while the SG ones $\nu_{SG}=1.22^{+0.26}_{-0.06}$ and $\eta_{SG}=-0.54^{+0.24}_{-0.52}$, where $\nu$  and $\eta$ are the correlation-length and the critical-point-decay exponents, respectively. These exponents turn out to be close to the corresponding Heisenberg SG exponents, but are different from the Ising SG ones. We also confirm the 1RSB nature of the ordered state, consistently with the previous reports.

This paper is organized as follows. In \Rsec{model}, we introduce the model and explain some of the details of our simulation. In \Rsec{quantity}, we define several physical quantities which we compute to examine the SG and the CG orderings. Our criteria for checking equilibration are also shown in this section. The results of our MC simulations are presented in \Rsec{result}. We show the data of the glass order parameter, the Binder parameter, the correlation-length ratio, the overlap-distribution function,  the non-self-averageness parameters and the spatial correlation function for both the SG and the CG degrees of freedom. The CG and the SG transition temperatures are estimated via an infinite-size extrapolation of appropriate finite-size data. Possible RSB character of the ordered state is also examined in this section. In  \Rsec{critical}, the critical properties of the SG and the CG transitions are analyzed, and the CG and the SG critical exponents are determined. Comparison is made with the corresponding exponents of the 3D Heisenberg and of the 3D Ising SGs. The last section is devoted to summary and discussion. In appendix, the behavior of the SG Binder parameter in the thermodynamic limit across the CG and the SG transition points is analyzed.


\section{\label{sec:model}MODEL AND SIMULATIONS}

The model we study is the isotropic {\it XY} EA model on a 3D simple-cubic lattice. The sites are labeled by the index $i$ ($i=1,2,\cdots,N$), the corresponding coordinate being denoted as $\V{r}_{i}=(x_{i},y_{i},z_{i})$. The total number of spins $N$ is related to the linear system size $L$ as $N=L^3$. The {\it XY\/} spin on the $i$th site, $\V{S}_i$, has two components, $\V{S}_i=(S_{ix},S_{iy})=(\cos\theta_i,\sin\theta_i)$ where $0 \leq \theta_{i} < 2 \pi$. The Hamiltonian is given by
\be
\mathcal{H}=-\sum_{\Ave{i,j}}J_{ij}\V{S}_{i}\cdot\V{S}_j,
\Leq{Hamiltonian}
\ee
where the summation $\Ave{i,j}$ is taken over the all nearest-neighbor pairs. The interaction $J_{ij}$ is a random Gaussian variable whose mean and variance are taken to be zero and unity, respectively. The partition function is given by
\be
Z=\int \prod_{i=1}^{N}\frac{d\theta_i}{2\pi} 
e^{-\beta \mathcal{H} },
\ee
where $\beta$ is the inverse temperature $1/T$ normalized by the Boltzmann constant $k_{\rm B}$. The thermal average will be denoted by the angular brackets $\Ave{\cdots}$.

 We perform MC simulations based on the single-spin-flip Metropolis method combined with the over-relaxation method and the temperature-exchange technique. This algorithm is known to effectively reduce the long correlation time involved in simulation of hard-relaxing systems such as SG.

 In a unit process of the over-relaxation, we compute first the local field felt by the spin at site $i$, $\V{h}_i=\sum_{j\in \Lambda_{i}} J_{ij}\V{S}_{j}$ where $\Lambda_{i}$ represents the neighbors of the site $i$, and then reflect the spin $\V{S}_i$ with respect to the local field $\V{h}_i$ as
\be
\V{S}_{i}\to \V{S}'_{i}=-\V{S}_{i}+2 \frac{\V{S}_i\cdot \V{h}_i}{\V{h}_i^2}\V{h}_i.
\ee

The simple cubic lattice consists of two interpenetrating sublattices, and we perform the Metropolis update sequentially through the sites on one sublattice after another, which is followed by the $M$-times over-relaxation sweeps, also performed sequentially through the sites on each sublattice. This procedure constitutes our unit MC step. In our simulations, we take $M$ equal to the linear system size $L$. 

In the temperature-exchange process, we prepare $N_{T}$ spin configurations at a set of temperatures distributed between $T_{\rm min}$ and $T_{\rm max}$. The maximum temperature $T_{\rm max}$ is chosen to be high enough where the autocorrelation times of the spin and the chirality are sufficiently short even in the single-spin-flip dynamics, typically 40 MC steps per spin (MCS). Each trial of the temperature exchange is performed for a pair of neighboring temperatures. A temperature-exchange trial is done after every MCS. 

In Table~\ref{tab-1}, we summarize the simulation parameters employed in our MC simulations, which include the linear system size $L$, the total number of averaged samples (independent bond realizations) $N_{s}$, the number of MCS discarded for equilibration $N_{MC1}$, the number of MCS employed for measuring physical quantities $N_{MC2}$, the maximum and the minimum temperatures in the temperature-exchange process $T_{\rm max}$ and $T_{\rm min}$,  and the number of temperature points $N_{T}$. Error bars are estimated  by using the bootstrap method from sample to sample fluctuations. 

\begin{table}[htbp]
\begin{center}
{\renewcommand\arraystretch{1.}
\begin{tabular}{lcccccc} 
\hline
$L$ & $N_s$ & $N_{MC1}$ & $N_{MC2}$ & $T_{\rm max}$ & $T_{\rm min}$ & $N_{T}$ \\ 
\hline
4   & 5000 & $1\times 10^4$   & $1\times 10^5$    & 0.86 & 0.24   &32 \\
6   & 5000 & $3\times 10^4$   & $1\times 10^5$    & 0.86 & 0.24   &32 \\
8   & 2000 & $5\times 10^4$   & $1\times 10^5$    & 0.80 & 0.24   &32 \\
12  & 2000 & $1\times 10^5$   & $1\times 10^5$    & 0.60 & 0.24   &32 \\
16  & 2048 & $4\times 10^5$   & $4\times 10^5$    & 0.52 & 0.26   &32 \\
20  & 1024 & $5\times 10^5$   & $5\times 10^5$    & 0.50 & 0.266  &40 \\
24  & 1024 & $7.5\times 10^5$ & $7.5\times 10^5$  & 0.49 & 0.271  &40 \\
32  & 1024 & $1.5\times 10^6$ & $1.5\times 10^6$  & 0.48 & 0.2736 &56 \\
40  & 384  & $2$-$2.8\times 10^6$   & $2$-$2.8\times 10^6$    & 0.46 & 0.2891 &64 \\
    & 128  & $2$-$3.4\times 10^6$   & $2$-$3.4\times 10^6$    & 0.442 & 0.2792 &64 \\
\hline
\end{tabular}
}
\end{center}
\caption{Parameters of our MC simulations. $L$ is the linear system size, $N_s$ is the total number of samples, $N_{MC1}$ is the number of MC steps per spin discarded for equilibration, $N_{MC2}$ is the number of  MC steps per spin subsequently used in measuring physical quantities, $T_{\rm max}$ and $T_{\rm min}$ are the highest and the lowest temperatures employed in the temperature-exchange process, and $N_{T}$ is the total number of temperature points. For $L=40$, we use two different temperature sets, and also we adaptively choose $N_{MC1}$ and $N_{MC2}$ for each bond realization to satisfy the equilibrium criteria shown below.}
\label{tab-1}
\end{table}
%


\section{PHYSICAL QUANTITIES}\Lsec{quantity}

In SGs, the conventional order parameter is an overlap between two independent systems with a common Hamiltonian. In the case of the {\it XY\/} model, each spin has two components and the spin overlap becomes a tensor with indices $\alpha$ and $\beta$ $(\alpha,\beta=x,y)$. We define the wavevector $\V{k}$-dependent spin overlap as
\be
q_{\alpha\beta}(\V{k})=\frac{1}{N}\sum_{i=1}^{N}S_{i\alpha}^{(1)}S_{i\beta}^{(2)}e^{i \V{k} \cdot \V{r}_{i} },
\ee
where the superscripts $(1)$ and $(2)$ denote two independent systems with the same Hamiltonian. For simplicity of notation, we write
\be
\hspace{-4mm}
q_{s}(\V{k})=\sqrt{
\sum_{\alpha,\beta}
\left| q_{\alpha\beta}(\V{k}) \right|^{2}}.
\ee

Similarly, we define the chirality and introduce the associated overlap. The chirality at a plaquette $p$ which is perpendicular to the $\mu(=x,y,z)$ axis is defined by
\be
\kappa_{p\perp \mu}=\frac{1}{2\sqrt{2}}\sum_{\Ave{i,j}\in p}\sgn{J_{ij}}\sin(\theta_i-\theta_{j}),
\ee
where the directed sum $\sum_{\Ave{i,j}\in p}$ is taken over four bonds surrounding the plaquette $p$ in a clockwise direction~\cite{Kawamura:01-1,Lee:03,Pixley:08}. The chiral overlap is then given by
\be
q_{\kappa}^{\mu}(\V{k})=\frac{1}{N}\sum_{p=1}^{N}\kappa_{p\perp \mu}^{(1)}\kappa_{p\perp \mu}^{(2)}e^{i\V{k}\cdot\V{r}_p}.
\ee

 The SG order parameter $q_{SG}$ and the SG susceptibility $\chi_{SG}$ are then defined by
\be
q_{SG}^{(2)}=\lsb \Ave{q_{s}(0)^2}\rsb, 
\hspace{1mm}
\chi_{SG}=Nq_{SG}^{(2)}, 
\ee
where the square brackets $\lsb \cdots \rsb$ denote the configurational average, {\it i.e.\/}, the average over the bond disorder.

 The SG Binder parameter $g_{SG}$ and the SG correlation length $\xi_{SG}$ based on the Ornstein-Zernike form are defined by
\be
&&g_{SG}=3-2
\frac{\lsb \Ave{ q_{s}(0)^4 } \rsb }
{\lsb \Ave{q_{s}(0)^2} \rsb^2},
\\
&&
\xi_{SG}=\frac{1}{2\sin\lb k_{\rm min}/2\rb}
\sqrt{
\frac{
\lsb \Ave{q_{s}(0)^2} \rsb
}{
\lsb \Ave{q_{s}(\V{k}_{\rm min})^2} \rsb
}-1},
\Leq{xi_SG}
\ee
where $\V{k}_{\rm min}=(2\pi/L,0,0)$. 

 On the other hand, the CG order parameter and the CG susceptibility are given by
\be
q_{CG}^{(2)}=\lsb \Ave{q_{\kappa}^{\mu}(0)^2} \rsb,
\hspace{1mm}
\chi_{CG}=N q_{CG}^{(2)}.
\ee
The direction $\mu$-dependence of the right-hand side should vanish after the sample average $\lsb \cdots \rsb$, and we take the average over $\mu=x,y,z$ in the actual calculation. The CG Binder parameter and the CG correlation length are  given by
\be
&&
g_{CG}=
\frac{1}{2}
\lb 3-
\frac{
\lsb \Ave{q_{\kappa}^{\mu}(0)^4} \rsb }
{\lsb \Ave{q_{\kappa}^{\mu}(0)^2} \rsb^2}
\rb,
\\
&&
\xi^{\mu}_{CG}=\frac{1}{2\sin\lb k_{\rm min}/2\rb}
\sqrt{
\frac{
\lsb \Ave{q_{\kappa}^{\mu}(0)^2} \rsb
}{
\lsb \Ave{
|q_{\kappa}^{\mu}(\V{k}_{\rm min})|^2} \rsb
}-1}.
\Leq{xi_CG}
\ee
Although the $\mu$-dependence again vanishes for $g_{CG}$, it remains for $\xi^{\mu}_{CG}$ due to the nontrivial wavevector dependence of $q_{\kappa}^{\mu}(\V{k}_{\rm min})$, {\it i.e.\/}, the dependence on the direction $\mu$ with respect to $\V{k}(\parallel \hat x$) taken here parallel with $\hat x$. We denote $\xi^{x}_{CG}$ as $\xi^{\parallel}_{CG}$ and $\xi^{y,z}_{CG}$ as $\xi^{\perp}_{CG}$, and will show both the data below.

We also define a parameter quantifying the non-self-averageness of the order parameter, the $A$ parameter~\cite{Marinari:99}. It is defined either for the spin or for the chirality by
\be
&&A_{SG}=
\frac{
\lsb \Ave{ q_{s}(0)^2 }^2 \rsb 
-
\lsb \Ave{q_{s}(0)^2 } \rsb^2
}
{
\lsb \Ave{q_{s}(0)^2 } \rsb^2
},
\\ 
&&
A_{CG}=
\frac{
\lsb \Ave{q_{\kappa}^{\mu}(0)^2}^2 \rsb 
-
\lsb \Ave{q_{\kappa}^{\mu}(0)^2} \rsb^2
}
{
\lsb \Ave{q_{\kappa}^{\mu}(0)^2} \rsb^2
}.
\ee
The $\mu$-dependence vanishes for $A_{CG}$. The $A$ parameter becomes nonzero if the SG or the CG susceptibility is non-self-averaging. Note that, even when $q_{SG}$ vanishes in the thermodynamic limit, $A_{SG}$ can become finite if $\chi_{SG}$ is non-self-averaging. In the current problem, such a situation can emerge in the temperature range $T_{SG}< T <T_{CG}$ in the possible occurrence of the spin-chirality decoupling.

We also introduce the so-called Guerra parameter or the $G$ parameter defined by
\be
&&G_{SG}=
\frac{
\lsb \Ave{q_{s}(0)^2}^2 \rsb 
-
\lsb \Ave{q_{s}(0)^2} \rsb^2
}
{
\lsb \Ave{q_{s}(0)^4} \rsb
-
\lsb \Ave{q_{s}(0)^2} \rsb^2
},
\\ 
&&
G_{CG}=
\frac{
\lsb \Ave{q_{\kappa}^{\mu}(0)^2}^2 \rsb 
-
\lsb \Ave{q_{\kappa}^{\mu}(0)^2} \rsb^2
}
{
\lsb \Ave{q_{\kappa}^{\mu}(0)^4} \rsb
-
\lsb \Ave{q_{\kappa}^{\mu}(0)^2} \rsb^2
}.
\ee
The $G$ parameter looks like the $A$ parameter, but there is a difference in that the $G$ parameter can be finite even when the ordered state does not accompany the RSB~\cite{Marinari:98,Bokil:99,Picco:01}.

 The distributions of the spin and the chiral overlaps might provide a signal of the RSB. In this paper, we examine the following two overlap distributions
\be
&&
P_{s}(q)=\lsb \delta\lb q-\sum_{\alpha}q_{\alpha\alpha}(0) \rb \rsb,
\\
&&
P_{\kappa}(q)=\lsb \delta\lb q-q_{\kappa}^{\mu}(0) \rb \rsb.
\ee
For the chiral overlap distribution $P_{\kappa}(q)$, the RSB effect is simple: if there is no RSB in the ordered state, the distribution has only two $\delta$-peaks in the thermodynamic limit which are related each other by the $Z_2$ spin-reflection symmetry of the whole spins. On the other hand, the spin overlap distribution $P_{s}(q)$ takes a non-trivial form even in the SG ordered state without the RSB, a superposition of two $\delta$-peaks located at $q=\pm q_{EA}$ and a broad distribution spanning between these diverging peaks, due to the projection of the tensor $q_{\alpha\beta}$ onto the diagonal component. This makes it rather difficult to obtain a clear indication of the RSB from the $P_{s}$ data. For further details, see Ref. 47.

Next we explain how we check the equilibration in our simulations. In any equilibrium simulations of SGs, a special care is required for the thermalization due to the hard relaxation of the system. Here we check the equilibration according to the following four criteria.

(1) All the temperature replicas should move back and forth many times along the temperature axis during the temperature-exchange process, while the relaxation at the highest temperature $T=T_{\rm max}$ is fast enough.  This criterion is  easy to implement, and is empirically known to be a stringent test of equilibration. We show in \Rfig{average-roundtrip}  the histogram of $N_{\rm round}$, the number of round trips between $T_{\rm min}$ and $T_{\rm max}$ averaged over all the temperature replicas,  for the simulated $N_{s}=512$ samples of our largest size $L=40$. One can see from the figure that all the samples satisfy the criteria of $N_{\rm round}\geq 5$ and most of the samples ($\sim 95\%$) satisfy the criterion of $N_{\rm round}\geq 10$. A harder criterion concerns with the minimum number of round trips among all the temperature replicas for each sample. We require that this number is greater than 3 for $99\%$ of samples. We check that, with this small fraction ($\sim 1\%$) of ``bad'' samples, either inclusion or exclusion of the bad samples does not change our final results for all the physical quantities.

\begin{figure}[htbp]
\includegraphics[width=1\columnwidth,height=0.7\columnwidth]{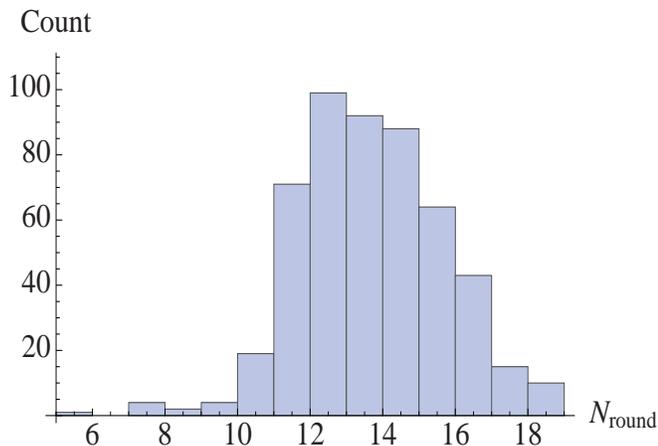}
\caption{\Lfig{average-roundtrip}(Color online) The histogram of the averaged number of round trips between $T_{\rm min}$ and $T_{\rm max}$ of the simulated $N_{s}=512$ samples for the largest size $L=40$.}
\end{figure}

(2) We check all the measured physical quantities converge to stable values. As an example, we show in \Rfig{MCS} the MC time dependence of the glass order parameter, the Binder parameter, and the correlation-length ratio both for the spin and the chiral degrees of freedom at $T_{\rm min}=0.2792$ for our largest system size $L=40$. The average is taken here over $N_{s}=128$ samples. 
\begin{figure}[htbp]
\includegraphics[width=1\columnwidth,height=0.65\columnwidth]{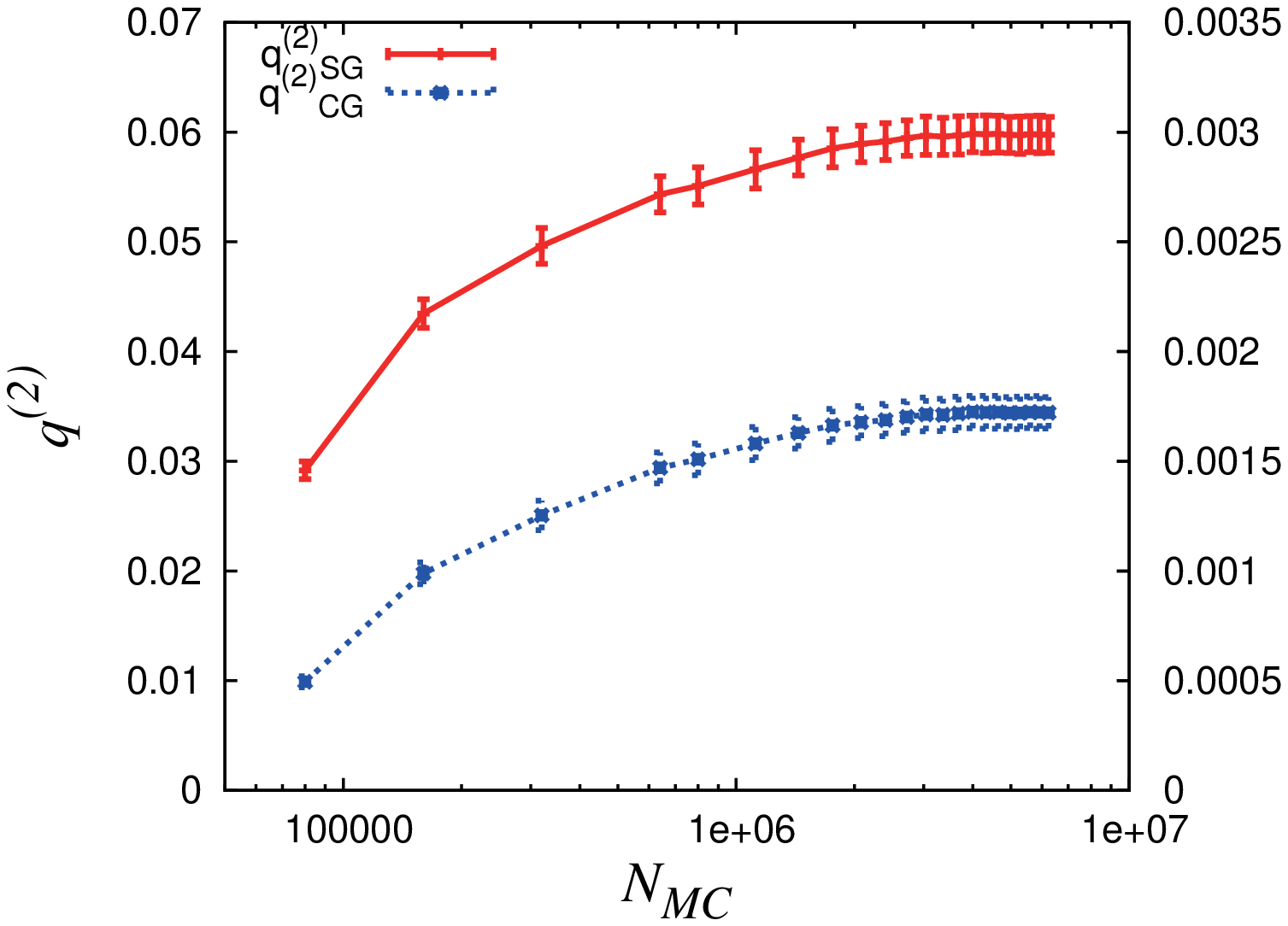}
\includegraphics[width=1\columnwidth,height=0.65\columnwidth]{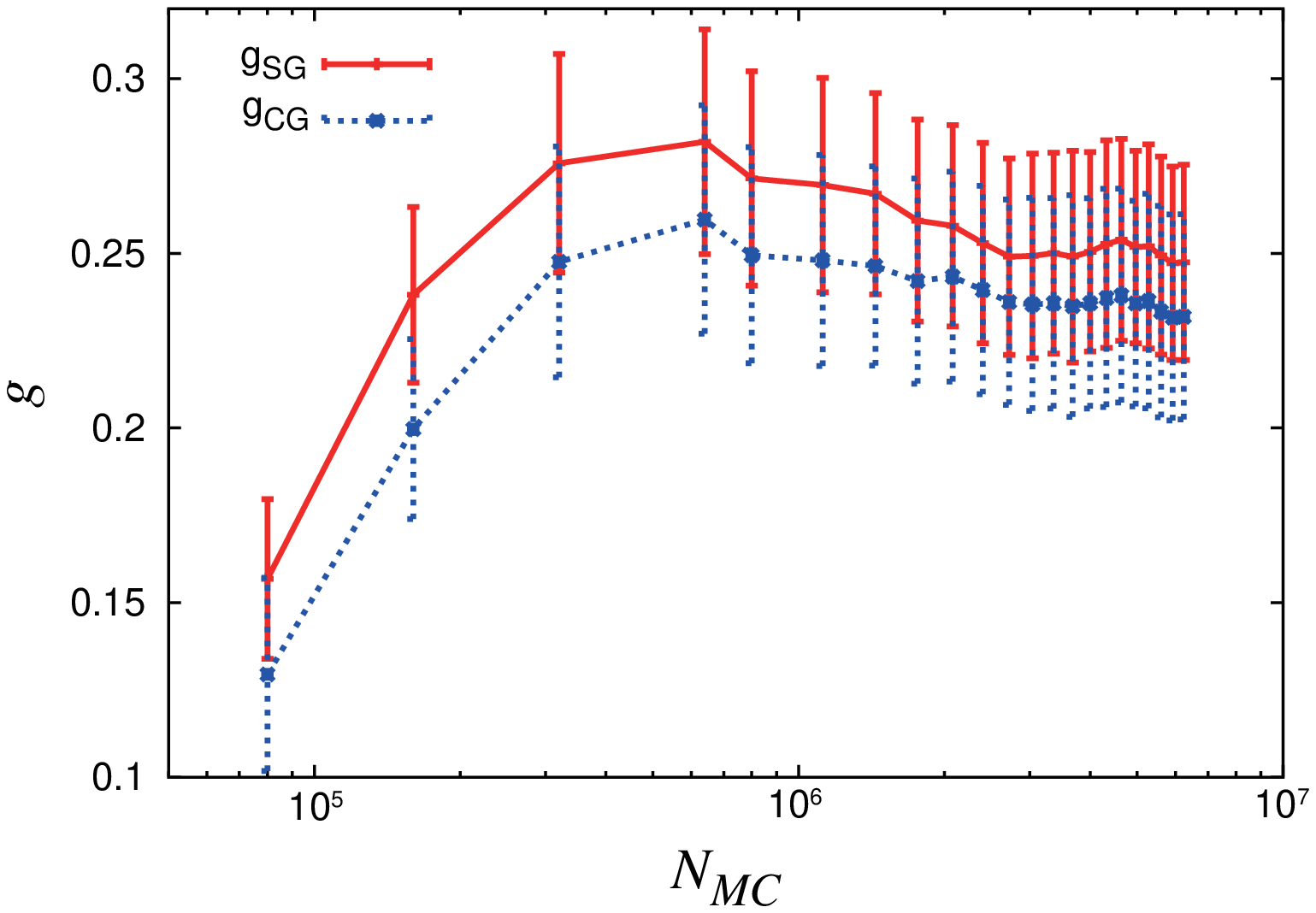}
\includegraphics[width=1\columnwidth,height=0.65\columnwidth]{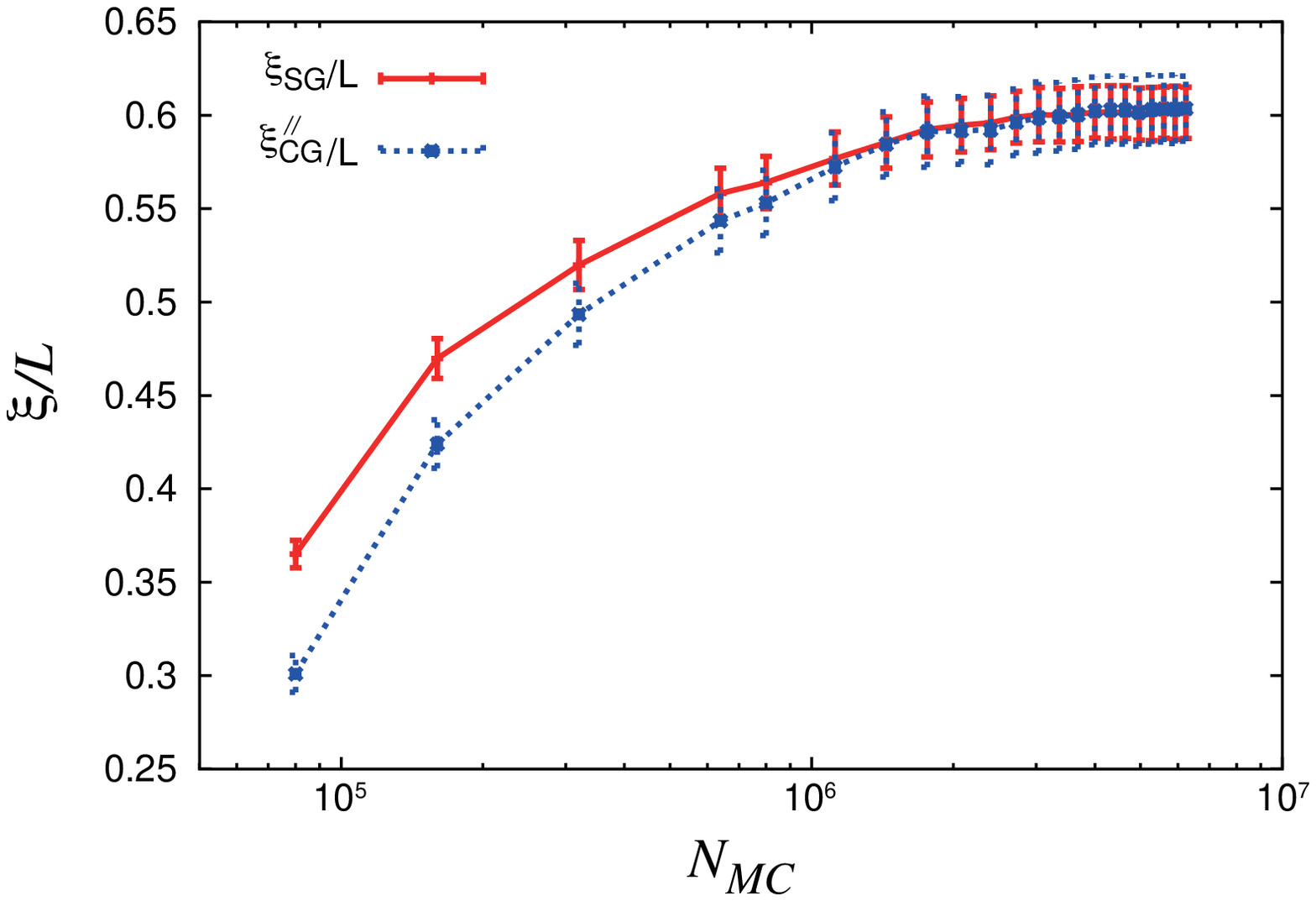}
\caption{\Lfig{MCS}(Color online) The MC time dependence of the glass order parameter (upper), the Binder parameter (middle), and the correlation-length ratio (lower) both for spin and chiral degrees of freedom where $N_{MC}$ represents the number of MCS. The temperature is $T_{\rm min}=0.2792$, and the lattice size is $L=40$, our largest system size. The average is taken over $128$ samples. In the upper figure, the left ordinate represents $q^{(2)}_{SG}$ while the right one represents $q^{(2)}_{CG}$.}
\end{figure}

(3) We confirm the equality between the specific heat computed via the energy fluctuation and the one computed via the temperature difference of the energy.

(4) The overlap-distribution functions $P_{s}(q)$ and $P_{\kappa}(q)$ should be  symmetric under the reversal operation $q \to -q$. We examine that this symmetry is satisfied {\it in each individual sample\/}. An example of $P_{s}(q)$ and $P_{\kappa}(q)$ for a typical $L=40$ sample is given in \Rfig{P-each} at a temperature $T=T_{\rm min}=0.2792$. (A clear three peak structure observed in $P_{\kappa}(q)$,  which is also observed in the averaged distribution function as shown later, is a 1RSB character.)  

\begin{figure}[htbp]
\includegraphics[width=0.7\columnwidth,height=1.0\columnwidth,angle=-90]{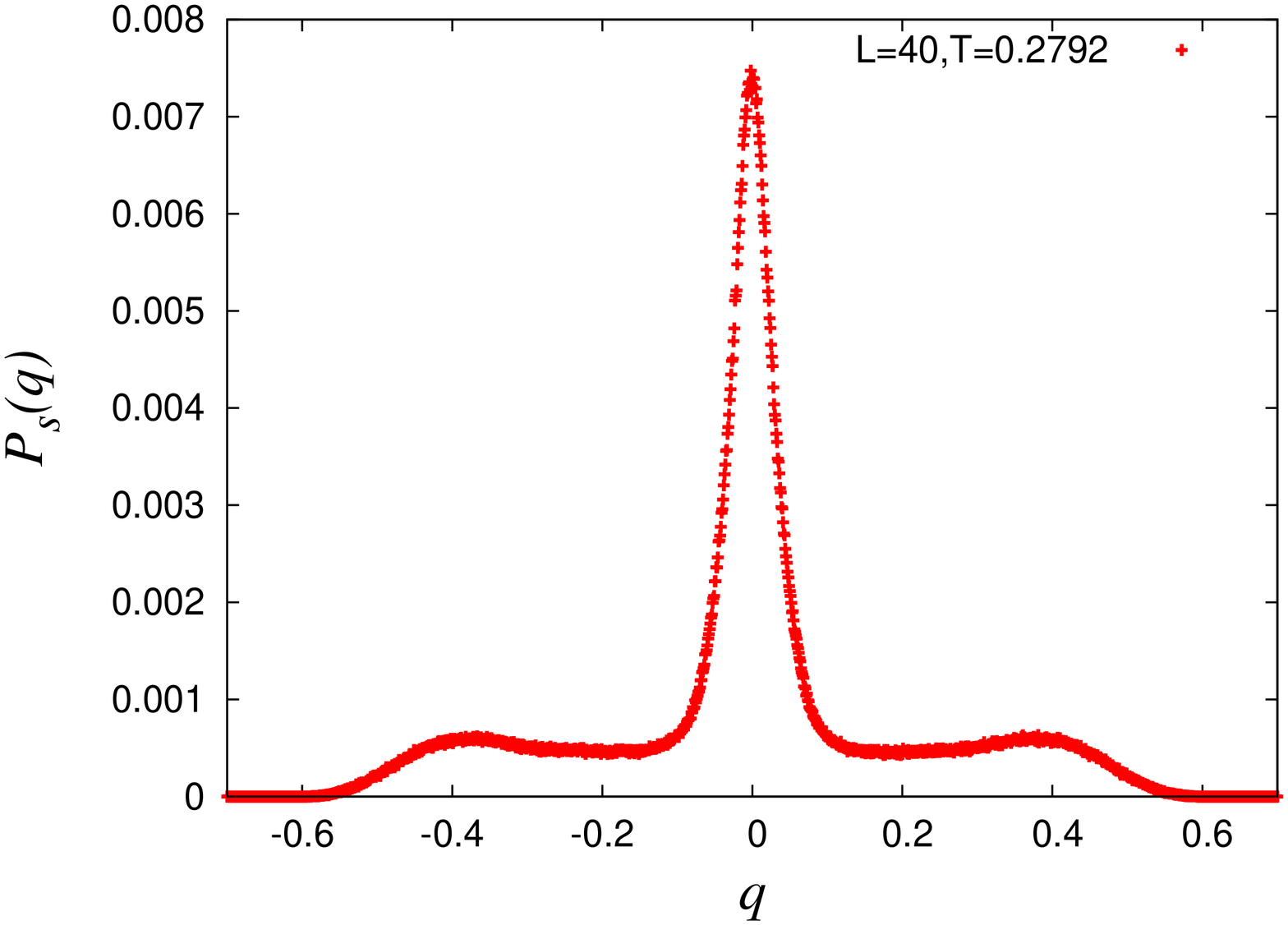}
\includegraphics[width=0.7\columnwidth,height=1.0\columnwidth,angle=-90]{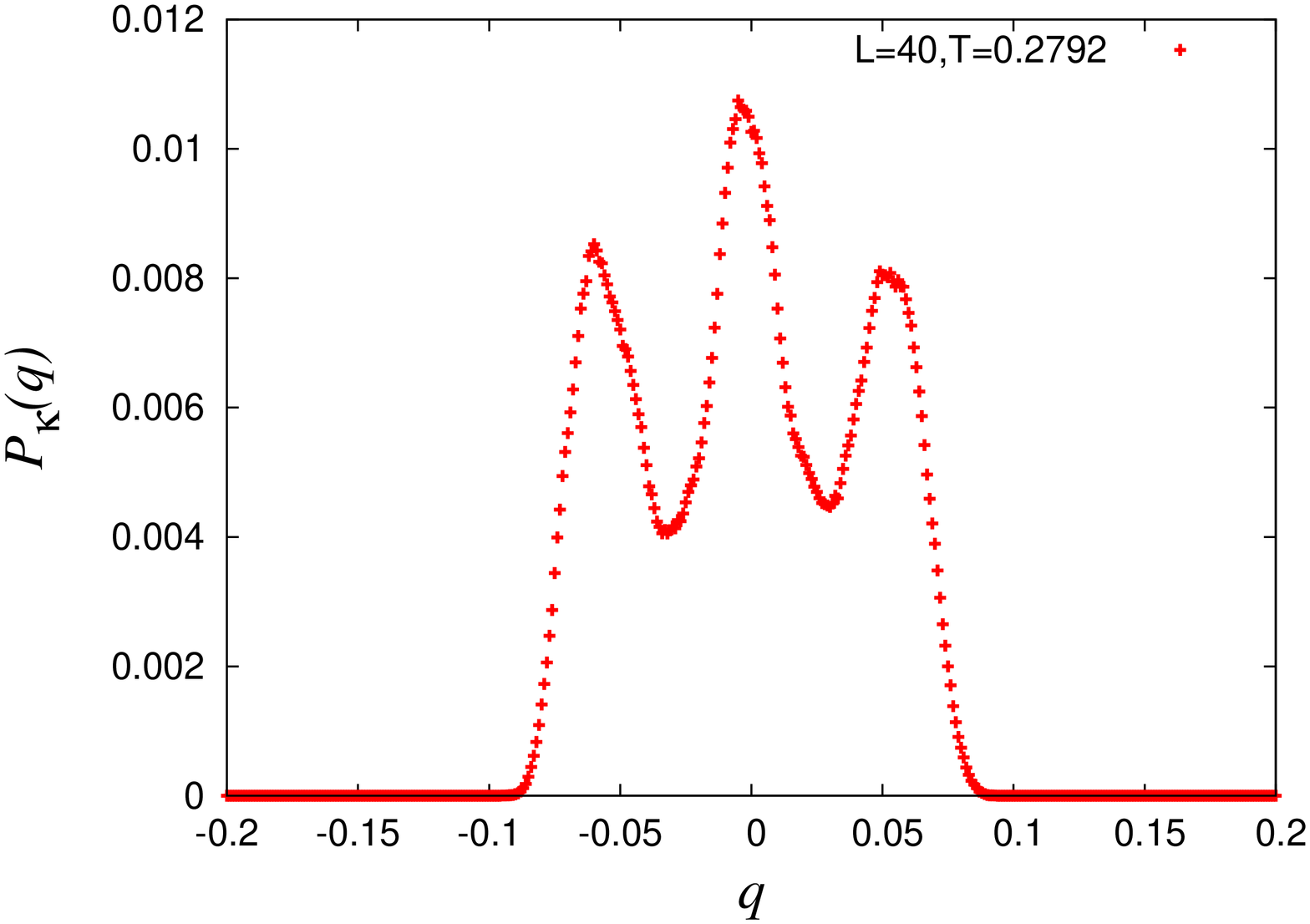}
\caption{\Lfig{P-each}(Color online) The spin (upper) and the chiral (lower) overlap-distribution functions $P_{s}(q)$ and $P_{\kappa}(q)$ for one particular bond realization (sample) of $L=40$ at the temperature $T=T_{\rm min}=0.2792$. The reversal symmetry $q \leftrightarrow -q$ is approximately satisfied. Three peak structure observed in $P_{\kappa}(q)$ is a characteristic of the 1RSB.}
\end{figure}

We think that these criteria (1)-(4) constitute sufficiently stringent tests of equilibration, and believe that the system is fully equilibrated up to the largest size $L=40$ and down to the lowest temperature $T=T_{\rm min}$ of our simulation.


\section{\Lsec{result}MONTE CARLO RESULTS}

In this section, we present the result of our MC simulations. 
We show first the temperature dependence of the specific heat in \Rfig{C}. No appreciable anomaly is seen in the specific heat, though the SG and the CG transition points actually exist in this temperature range, as will be shown below. The CG and the SG critical temperatures $T_{CG}$ and $T_{SG}$ are denoted by arrows in the figure.
\begin{figure}[htbp]
\includegraphics[width=1\columnwidth,height=0.7\columnwidth]{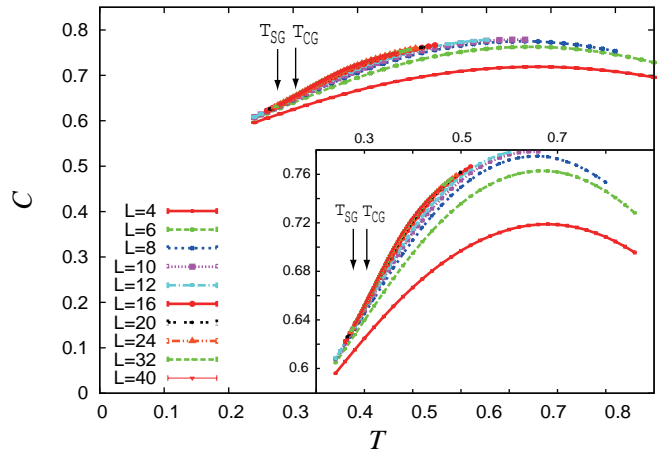}
\caption{\Lfig{C}(Color online) The temperature and size dependence of the specific heat per spin. Magnified view is given in the inset. The arrows at higher and lower temperatures indicate the CG and the SG transition points, respectively.
}
\end{figure}

 To investigate the SG and the CG orderings, we show the temperature dependence of the SG and the CG susceptibilities, $\chi_{SG}$ and $\chi_{CG}$, in \Rfig{chi}. In contrast to the SG susceptibility $\chi_{SG}$, which tends to increase as the system size $L$ is increased in the whole temperature region, the CG susceptibility $\chi_{CG}$ shows such a behavior only in the temperature region $T\lsim 0.4$, whereas, in the region $T\gsim 0.4$, it exhibits an opposite size dependence. This implies that the CG critical region is relatively narrow, which is common with the observation of the 3D Heisenberg SG~\cite{Viet:09}.
\begin{figure*}[htbp]
\includegraphics[width=1\columnwidth,height=0.75\columnwidth]{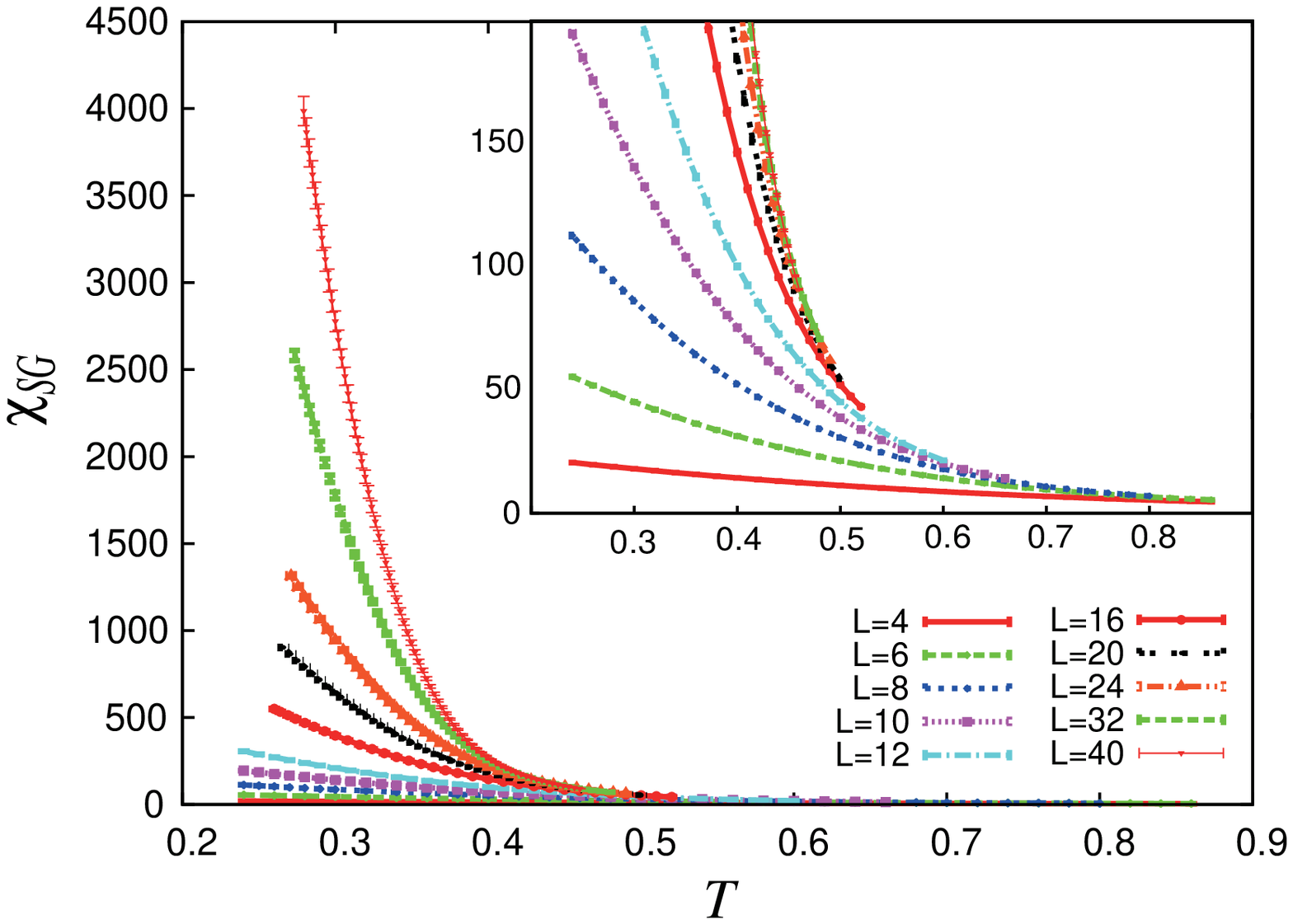}
\includegraphics[width=1\columnwidth,height=0.75\columnwidth]{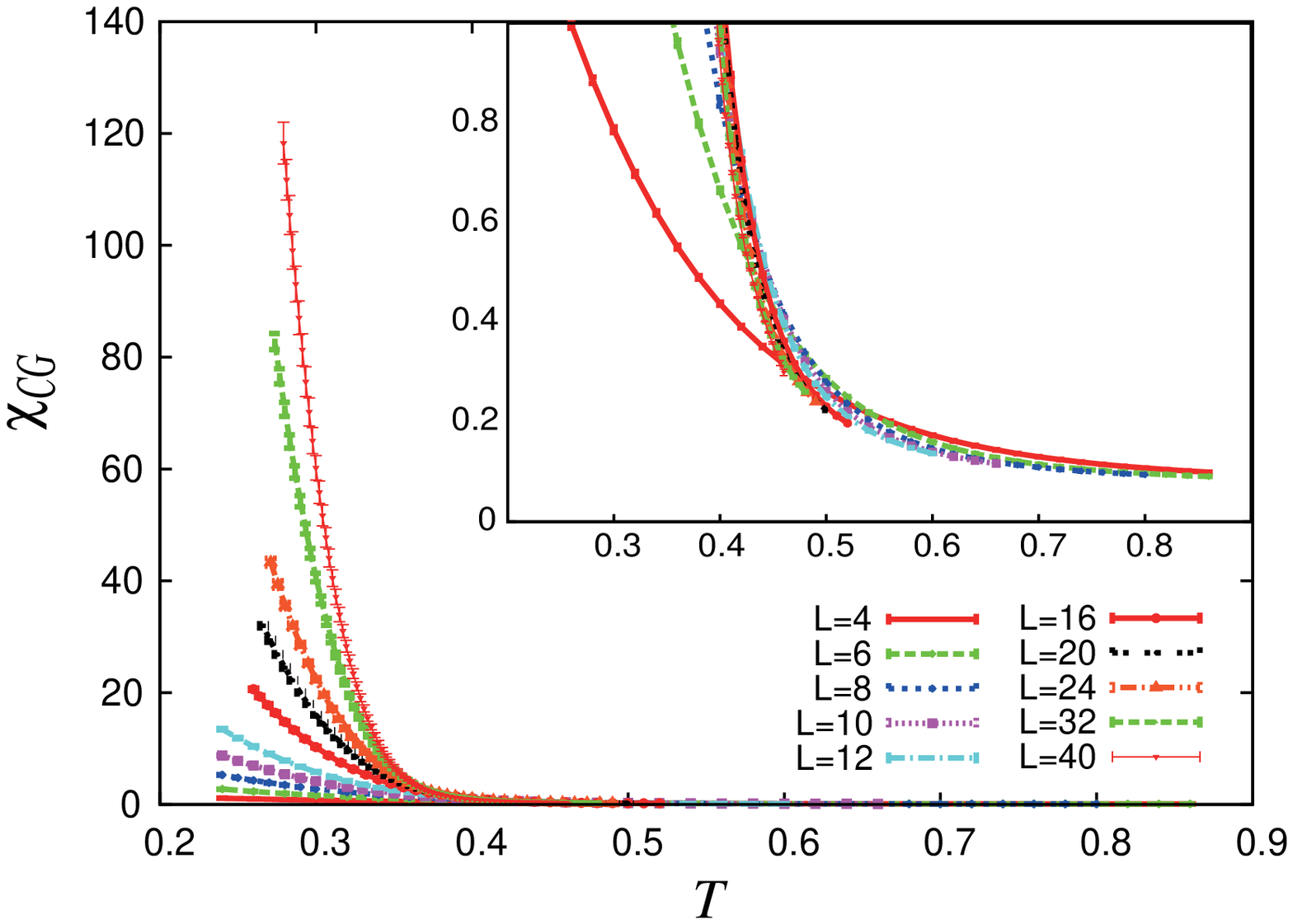}
\caption{\Lfig{chi}(Color online) The temperature and size dependence of the SG susceptibility $\chi_{SG}$ (left), and of the CG susceptibility $\chi_{CG}$ (right). The insets are magnified views. As can be seen from the insets, the magnitude of $\chi_{SG}$ increases as the system size is increased in the whole temperature region, while that of $\chi_{CG}$ exhibits an opposite size dependence in the temperature region $T\gtrsim 0.4$.}
\end{figure*}

 In \Rfig{L-q}, we plot the size dependence of the SG and the CG order parameters $q_{SG}^{(2)}$ and $q_{CG}^{(2)}$  for several temperatures on a double-logarithmic plot.
\begin{figure*}[htbp]
\includegraphics[width=1\columnwidth,height=0.7\columnwidth]{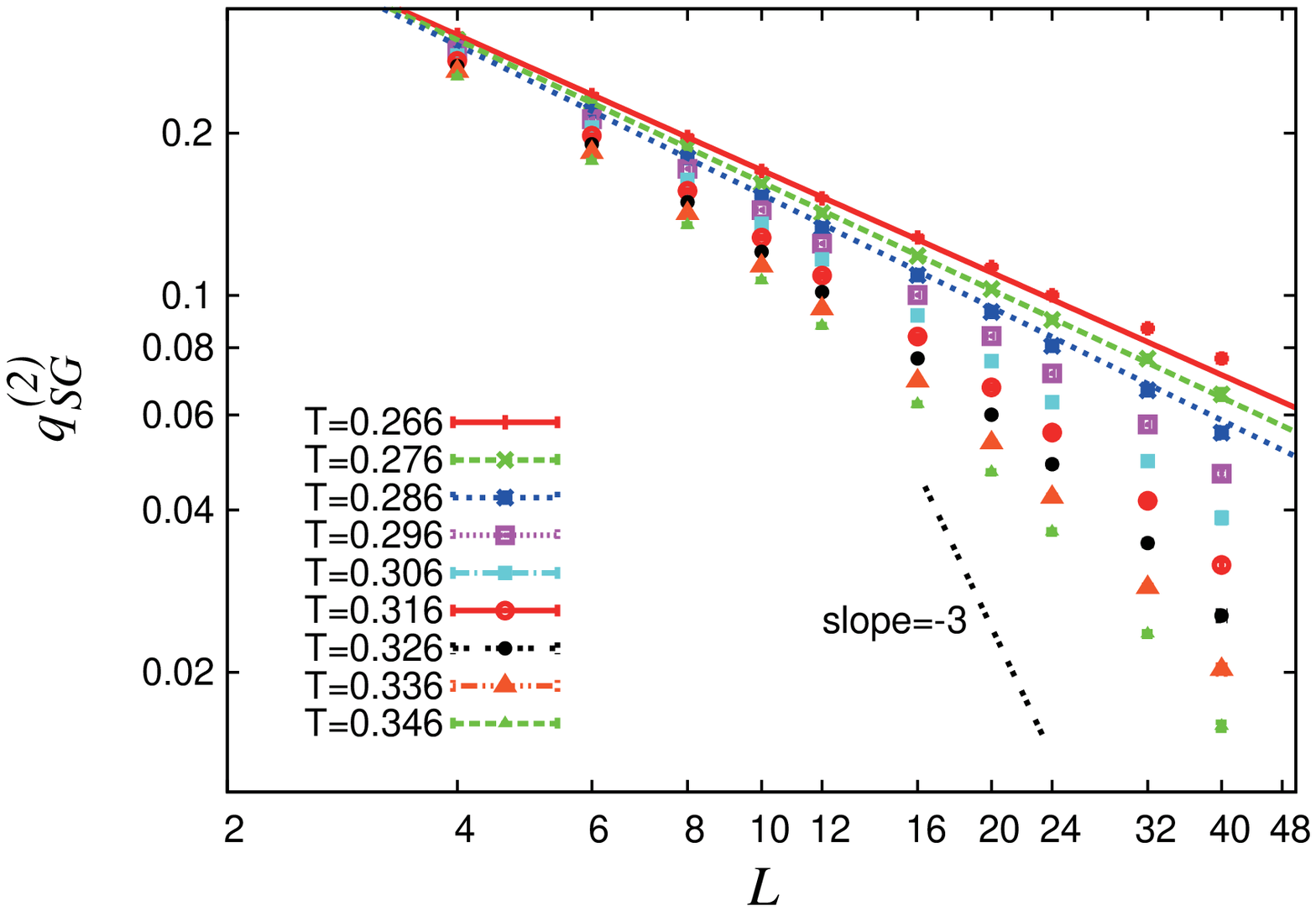}
\includegraphics[width=1\columnwidth,height=0.7\columnwidth]{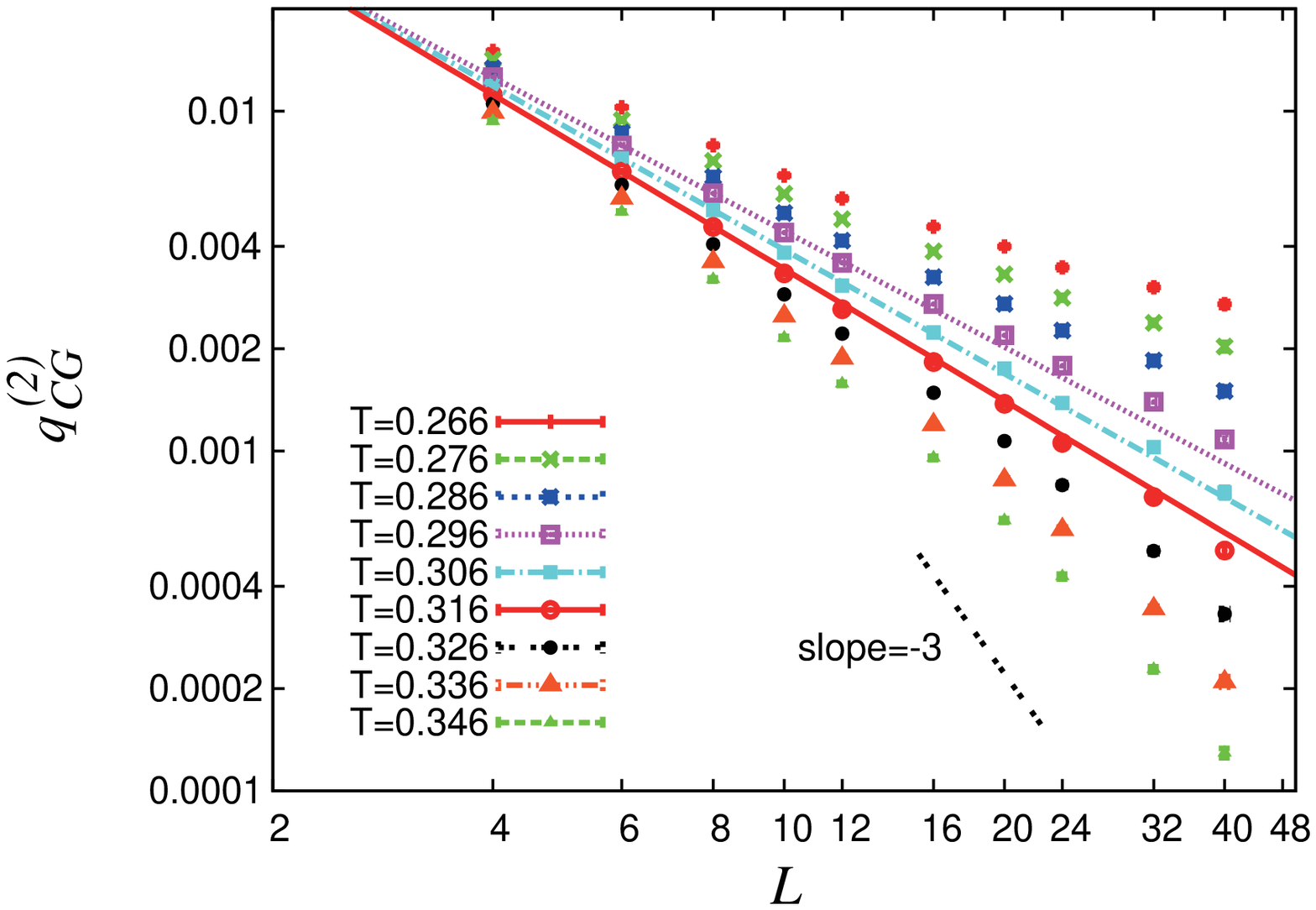}
\caption{\Lfig{L-q}(Color online) The size dependence of the SG order parameter (left), and of the CG order parameter (right) on a log-log plot for several temperatures. Straight lines are drawn by fitting the three data points of smaller sizes of $L=4,6,8$. The  $L=24,32$ data for $T=0.266$ as well as the $L=40$ data for $T=0.266$ and $0.276$ are obtained by extrapolating the higher temperature data to lower temperatures. A line with a slope $-d=-3$, which is the expected large-$L$ asymptotic behavior in the disordered phase, is also drawn.}
\end{figure*}
The data of the CG order parameter $q_{CG}^{(2)}$ exhibit a straight-line behavior around a temperature $T \sim 0.306$. It exhibits a clear upward trend  at lower temperatures implying the appearance of the CG long-range order, whereas at higher temperatures it exhibits a downward trend, eventually approaching another straight line with a slope $-d=-3$ generally expected  in the disordered phase for large enough systems. Then, we get our first estimate of the CG transition temperature,  $T_{CG}\sim 0.31\pm 0.015$.  By contrast, $q_{SG}^{(2)}$ exhibits such an upward trend only at the lowest temperature studied, $T=0.266$, with a straight-line behavior observed around $T \sim 0.276$. Then, we get an estimate of the SG transition temperature, $T_{SG}= 0.28\pm 0.015$. Hence, our data of the size dependence of the glass order parameters $q^{(2)}$ suggest that the spin-chirality decoupling really occurs in the present model. 

 In \Rfig{g}, we show the SG and the CG Binder parameters.
\begin{figure*}[htbp]
\includegraphics[width=1\columnwidth,height=0.75\columnwidth]{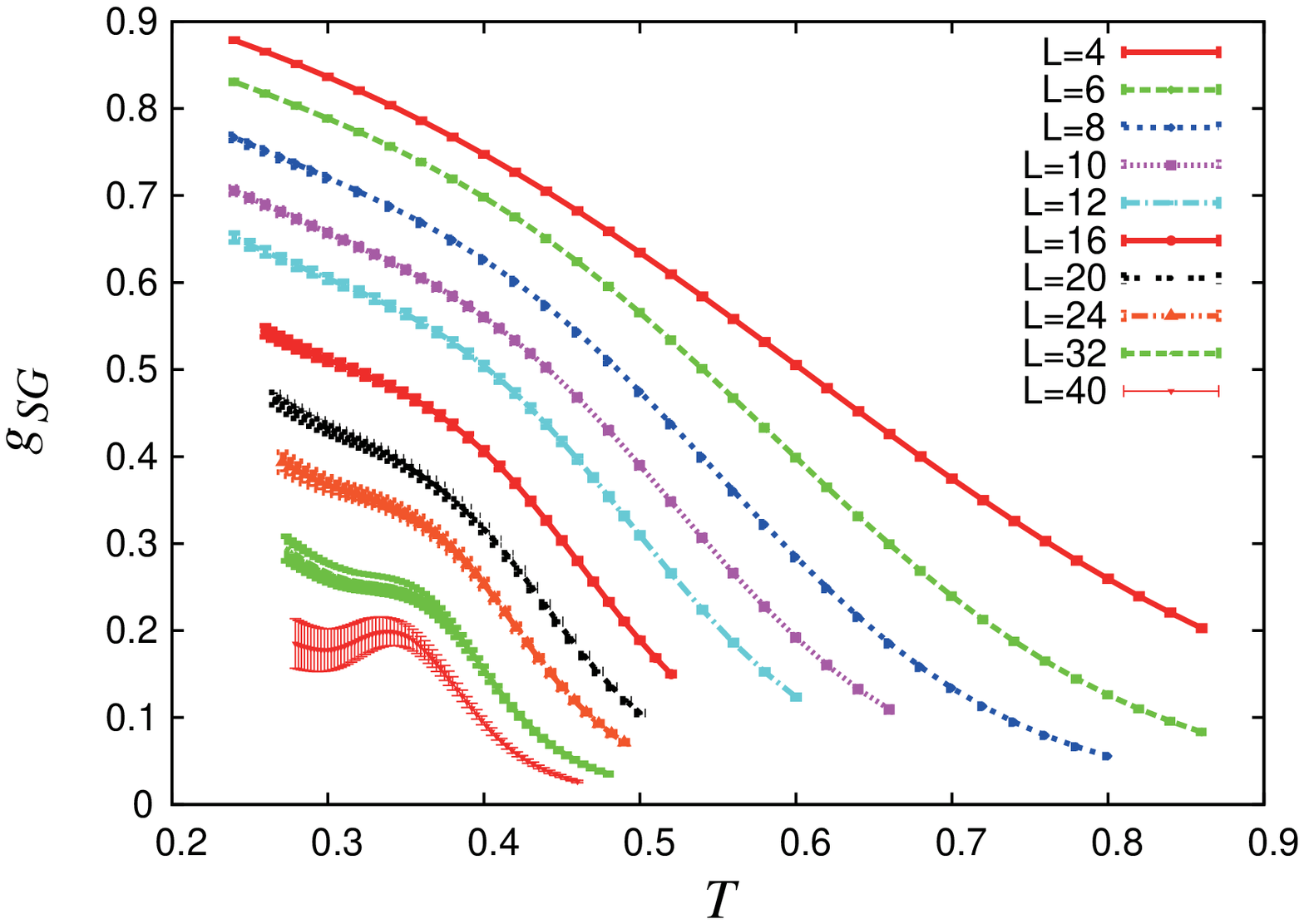}
\includegraphics[width=1\columnwidth,height=0.75\columnwidth]{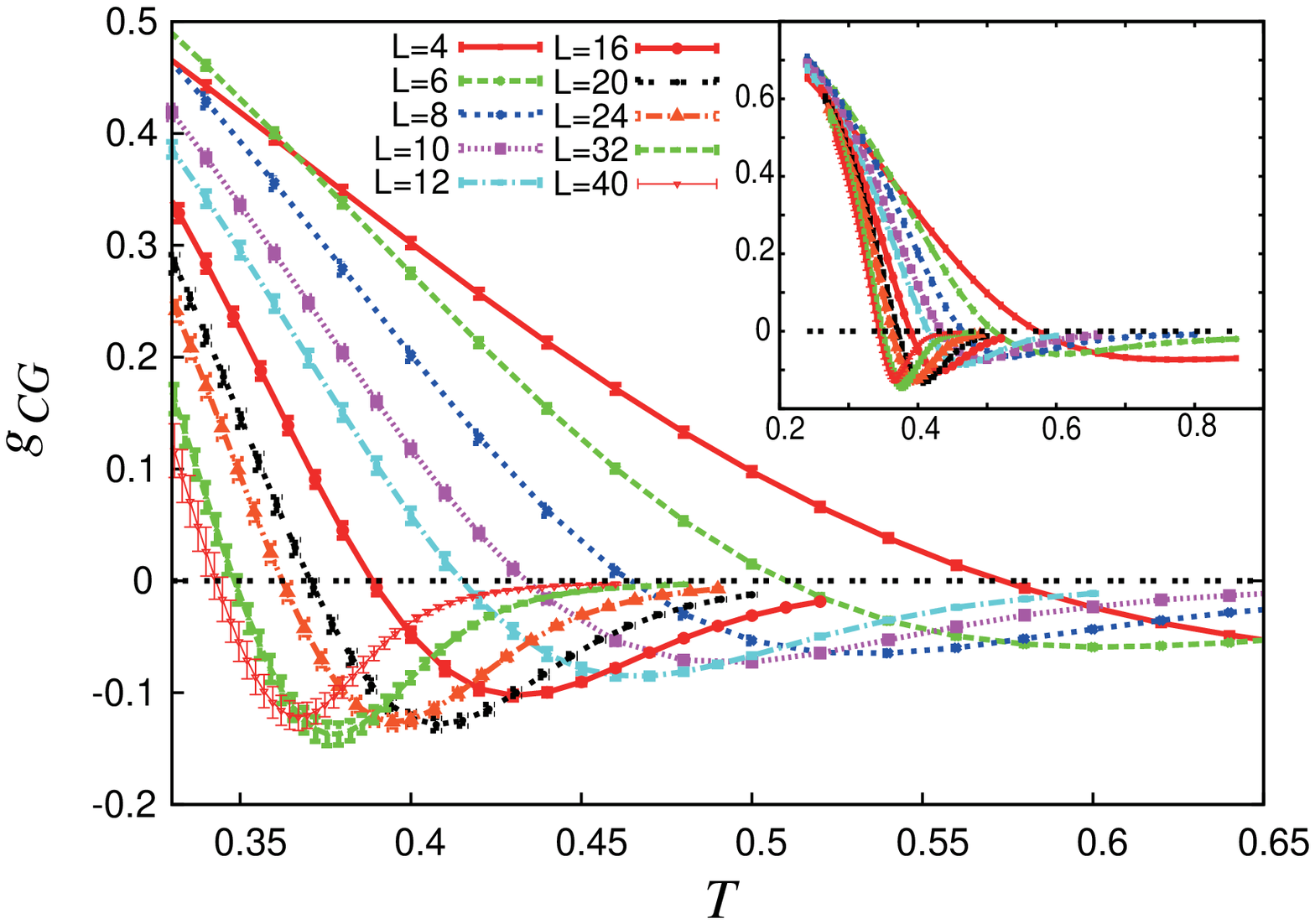}
\caption{\Lfig{g}(Color online) The temperature and size dependence of the SG Binder parameter $g_{SG}$ (left), and of the CG Binder parameter $g_{CG}$ (right). The inset of the right panel exhibits $g_{CG}$ in the wider temperature range.}
\end{figure*}
Consistently with the earlier reports~\cite{Kawamura:95-1,Kawamura:01-1}, the CG Binder parameter exhibits a non-divergent dip and a crossing among different sizes on the negative side of $g_{CG}$. Such a behavior is expected in a system exhibiting a continuous one-step RSB (1RSB). The crossing and the dip temperatures are expected to converge to $T_{CG}$ in the thermodynamic limit, which might provide a way to precisely estimate $T_{CG}$. By contrast, the SG Binder parameter $g_{SG}$ shows no crossing nor dip, decreasing monotonically as the system size is increased, though some characteristic inflections can be observed in a subtle way at large-size systems. So far, information concerning the transition points has been hard to obtain from $g_{SG}$. We re-examine the relation of $g_{SG}$ to the CG and the SG transition points below in appendix, and point out a possibility to extract information about $T_{SG}$ and $T_{CG}$ from the $g_{SG}$ data. 

 In \Rfig{gziSG}, we show the temperature dependence of the SG correlation-length ratio $\xi_{SG}/L$. Those of the parallel CG  correlation-length ratio $\xi_{CG}^{\parallel}/L$ and of the perpendicular CG  correlation-length ratio $\xi_{CG}^{\perp}/L$ are given in \Rfig{gziCG}.
\begin{figure}[htbp]
\includegraphics[width=1\columnwidth,height=0.75\columnwidth]{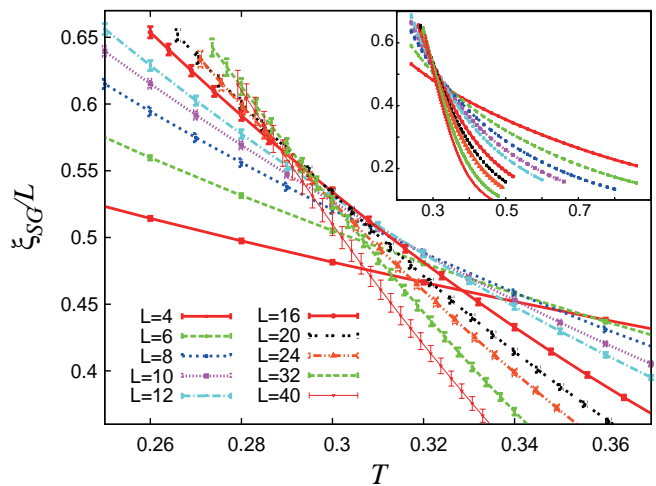}
\caption{\Lfig{gziSG}(Color online) The temperature and size dependence of the SG correlation-length ratio $\xi_{SG}/L$. The inset exhibits the wider temperature range.}
\end{figure}
\begin{figure*}[htbp]
\includegraphics[width=1\columnwidth,height=0.75\columnwidth]{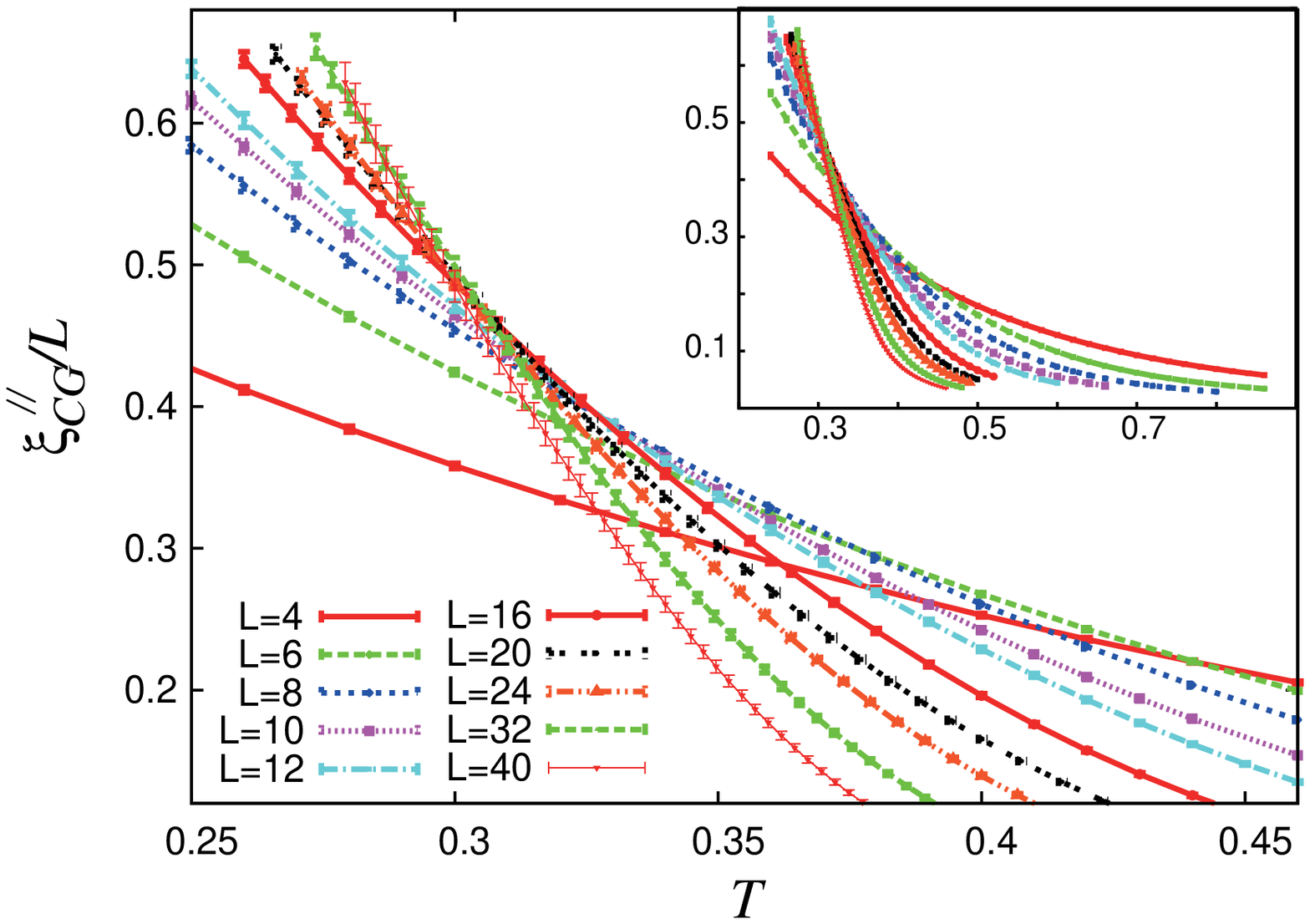}
\includegraphics[width=1\columnwidth,height=0.75\columnwidth]{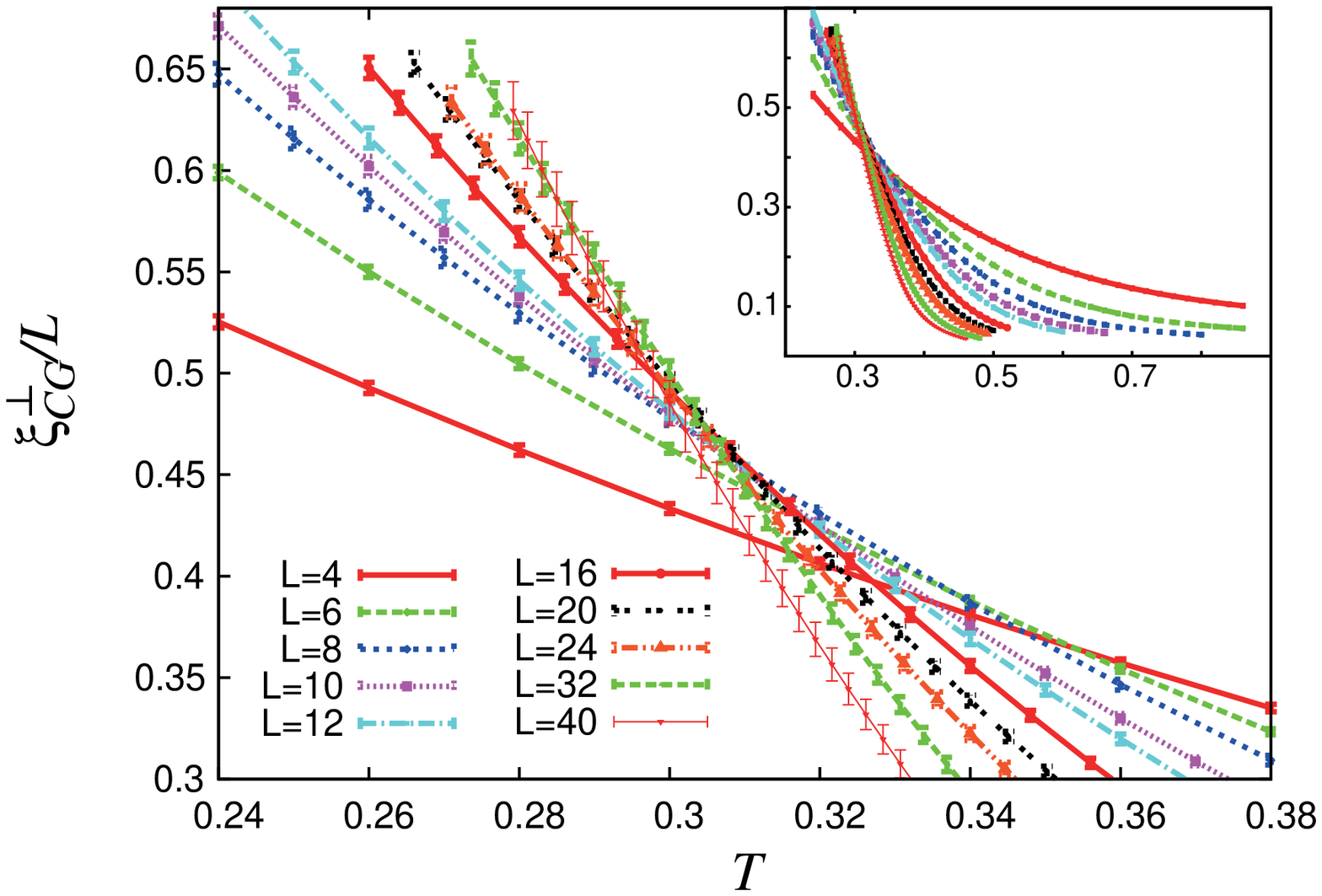}
\caption{\Lfig{gziCG}(Color online) The temperature and size dependence of the CG parallel correlation-length ratio $\xi_{CG}^{\parallel}/L$ (left), and of  the CG perpendicular correlation-length ratio $\xi_{CG}^{\perp}/L$ (right). The insets exhibit the wider temperature range.}
\end{figure*}
Both the SG and the CG correlation-length ratios show crossing among different sizes, and the crossing temperatures are expected to converge to the corresponding critical temperatures in the $L\rightarrow \infty$ limit. 

To estimate $T_{SG}$, we plot in the left panel of \Rfig{Tc} the crossing temperatures of the SG correlation-length ratio $\xi_{SG}/L$  for pairs of sizes $L$ and $sL$ with $s=2,3/2,4/3,5/3$ and $5/4$ versus $1/L_{ave}$ where $L_{ave}=(1+s)L/2$. Note that the crossing temperature for the pair $(L,sL)=(32,40)$ with $s=5/4$ is estimated by extrapolating the data to lower temperatures, since the raw data of $\xi_{SG}/L$ do not show a crossing in the investigated temperature range.

\begin{figure*}[htbp]
\includegraphics[width=1\columnwidth,height=0.75\columnwidth]{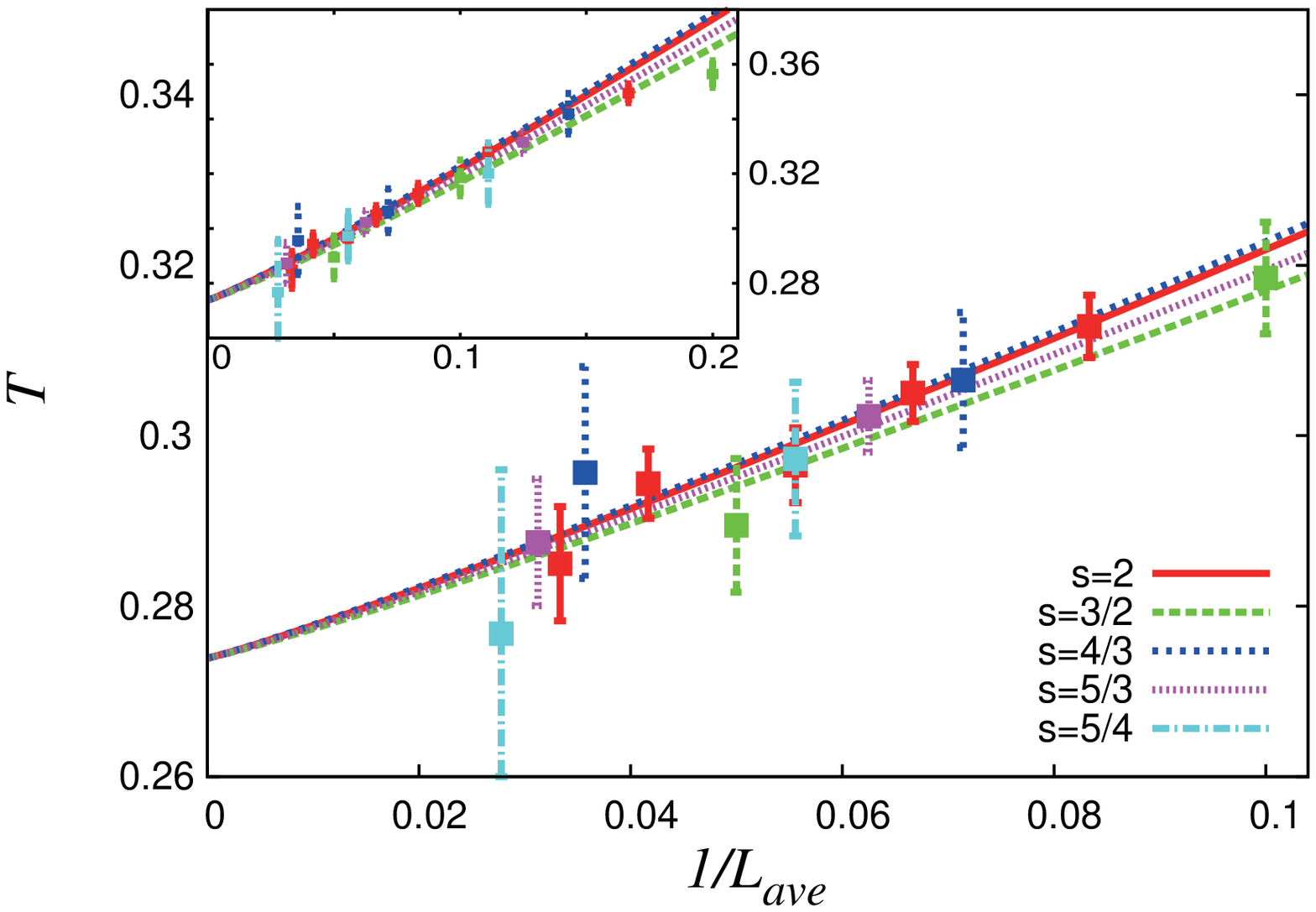}
\includegraphics[width=1\columnwidth,height=0.75\columnwidth]{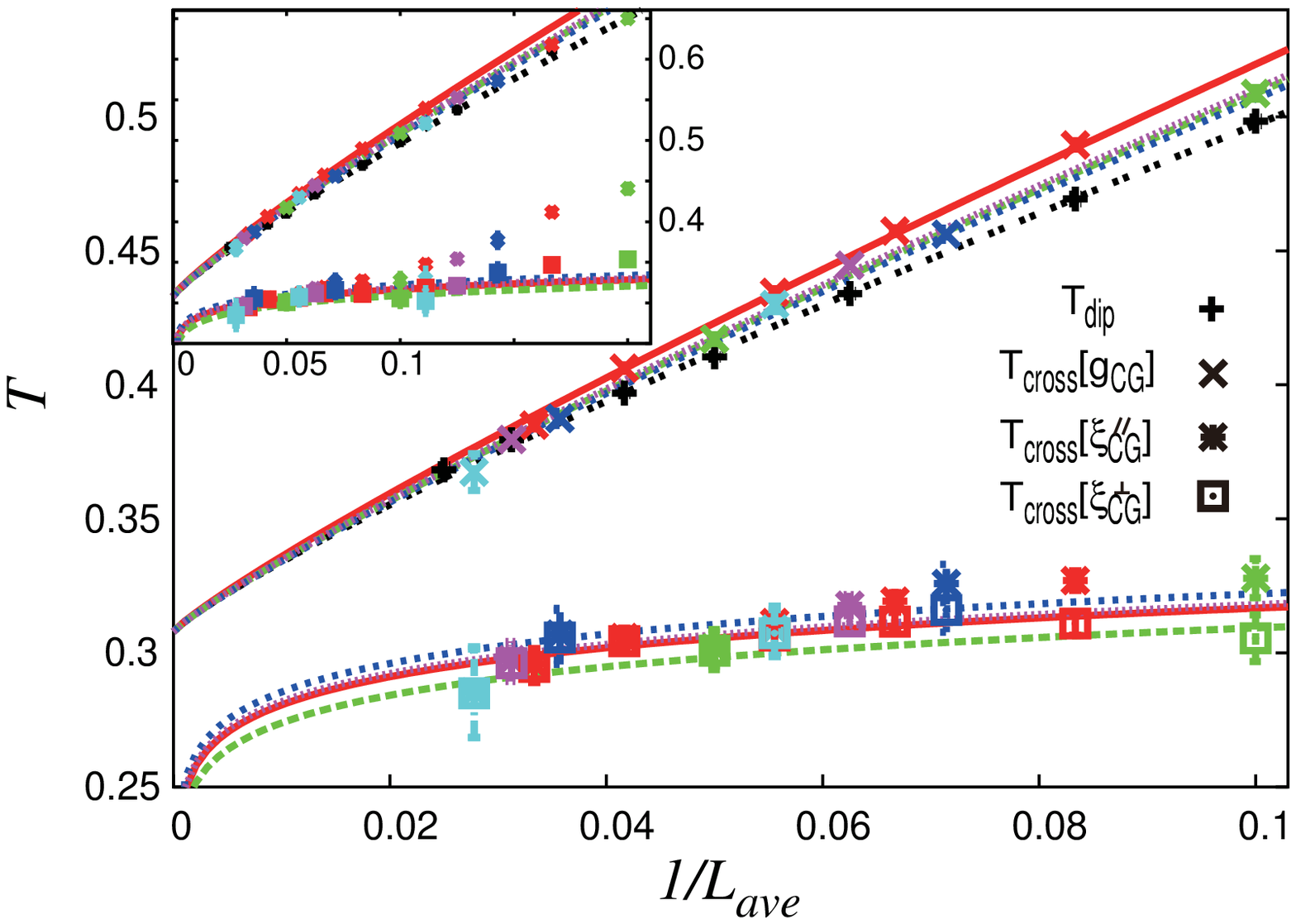}
\caption{\Lfig{Tc}(Color online) The crossing temperatures $T_{cross}$ of the SG correlation-length ratio $\xi_{SG}/L$ of the two sizes  $L$ and $sL$ are plotted versus the inverse mean lattice size $1/L_{ave}$ where $L_{ave}=(1+s)L/2$ (left). The crossing temperatures of the CG perpendicular correlation-length ratio $\xi_{CG}^\perp/L$ and of the CG Binder parameter $g_{CG}$ as well as the dip temperature $T_{dip}$ of $g_{CG}$ are plotted versus $1/L_{ave}$ (or $1/L$) (right). The insets exhibit the wider size range.}
\end{figure*}

We then try an infinite-size extrapolation of the crossing temperature $T_{cross}(L)$ based on the scaling from,

\be
T_{cross}(L)=T_{c} + cL^{-\theta}, \ \ \ \theta=\omega+\frac{1}{\nu}.
\Leq{T_cross}
\ee

In the fit of $T_{cross}(L;s)$ of the SG correlation-length ratios $\xi_{SG}/L$, we perform a combined fit of different $s$-sequences with a common $T_c=T_{SG}$ and a common $\theta=\theta_{SG}$. We use in the fit the data for $s=2,3/2,4/3$ and $5/3$ only, not the one for $s=5/4$ because of a possible inaccuracy due to the extrapolation employed for the data of $(L,sL)=(32,40)$ mentioned above. The resulting fitting curves are shown in the left panel of \Rfig{Tc}. The optimal fit is obtained for $T_{SG}=0.274$ and $\theta_{SG}=1.1$. If all data points were to be independent, which is actually not the case, one would get the associated error bars as $T_{SG}=0.274^{+0.012}_{-0.032}$ and $\theta_{SG}=1.10^{+0.60}_{-0.55}$, where the error bar is underestimated. In order to get a more sensible estimate of the error bar, we use the $s=2$ series only, to get  $T_{SG}=0.275^{+0.013}_{-0.052}$ and $\theta_{SG}=1.14^{+0.75}_{-0.71}$. One sees that the use of merely the $s=2$ series changes the result only slightly. In the inset of  \Rfig{chi2-both}, we show the total $\chi$-square value of this fit with using the $s=2$ series versus the assumed $T_{SG}$ value. The horizontal line in the figure represents the total $\chi$-square value greater than the optimal value observed at $T_{SG}=0.275$ by unity, which gives our error criterion. The asymmetry of the curve results in the different values for the upper and the lower error values. We then finally quote $T_{SG}=0.275^{+0.013}_{-0.052}$ and $\theta_{SG}=1.1^{+0.75}_{-0.71}$.

\begin{figure}[htbp]
\includegraphics[width=1\columnwidth,height=0.8\columnwidth]{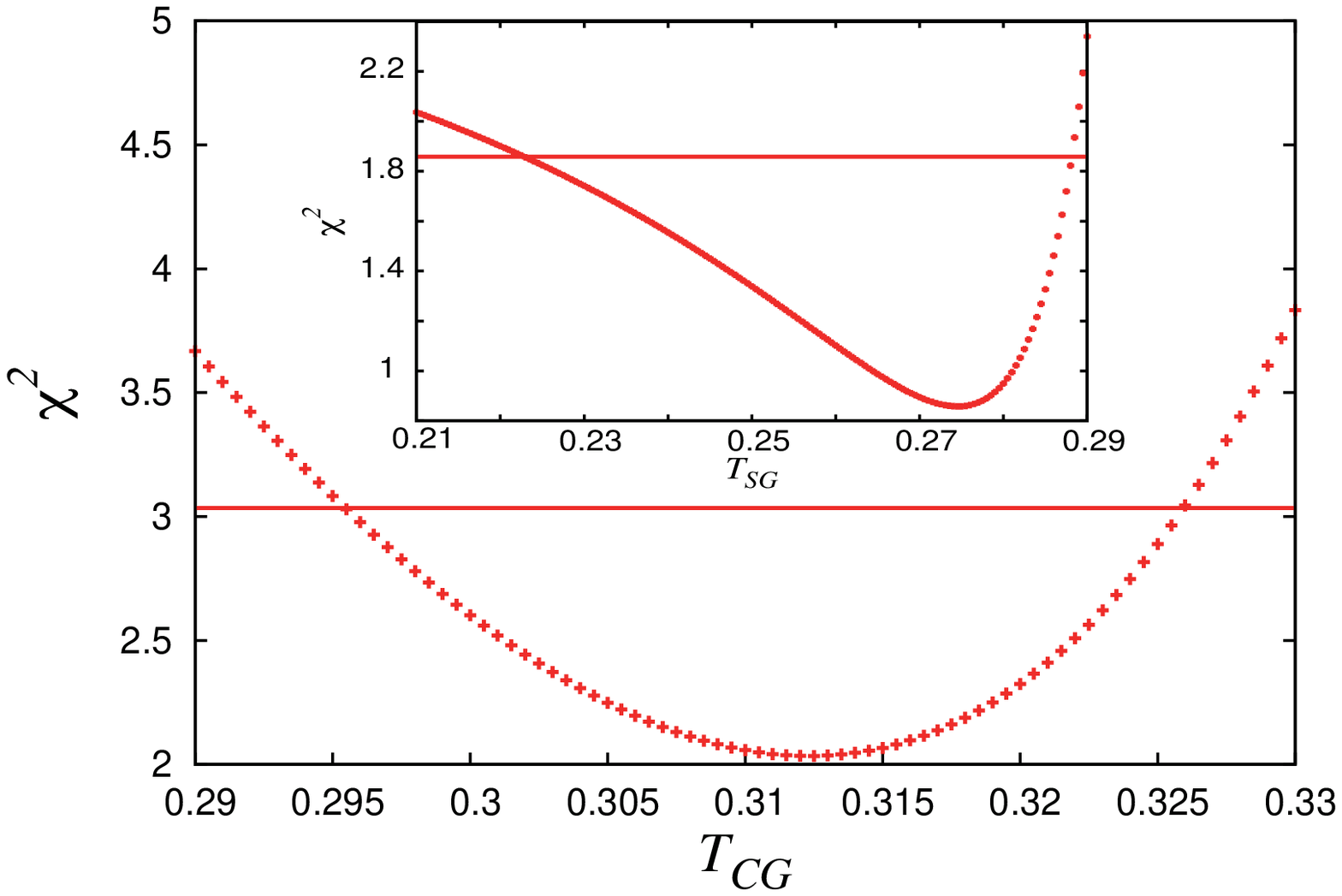}
\caption{\Lfig{chi2-both}(Color online) The total $\chi$-square values associated with the combined fit of the crossing and dip temperatures of the CG binder parameters $g_{CG}$ are plotted versus the CG transition temperature $T_{CG}$ assumed in the fit. The horizontal line represents the total $\chi$-square value greater than the optimal value by unity, usually used as an error criterion. The corresponding plot of the total $\chi$-square values obtained in our estimate of $T_{SG}$ is also presented in the inset.}
\end{figure}

 Next we turn to the estimate of $T_{CG}$ based on the CG correlation-length ratio $\xi_{CG}/L$. Although the parallel one $\xi^{\parallel}_{CG}/L$ and the perpendicular one $\xi^{\perp}_{CG}/L$ give two different sequences of the crossing temperatures, they tend to accord for $L_{ave} \gsim 8$ as can be seen in the right panel of \Rfig{Tc} (see also the inset). Hence,  to extract $T_{CG}$, we use the data of $\xi^{\perp}_{CG}$ for $L_{ave} \geq 8$ only. Unfortunately, and somewhat unexpectedly, the result of the fit of  $\xi_{CG}/L$ turns out to be rather pathological. The estimated best $T_{CG}$-value becomes extremely small or even negative. The resultant fitting curves are shown in the right panel of \Rfig{Tc} (the lower curves). We also examine the possible change in the fit with varying the lowest size used in the fit, but the pathology cannot be cured. The cause of this pathology is not entirely clear, but may be due to the peculiar behavior of the chiral-glass correlations. For example, $\chi_{CG}$ shows a non-monotonic size dependence absent in the corresponding $\chi_{SG}$. To further clarify this point, we discuss the properties of the spatial CG correlation function below in this section. Anyway,  in the present analysis, we abandon the crossing temperatures of $\xi_{CG}/L$ in our estimate of $T_{CG}$.

 In estimating $T_{CG}$, the crossing and the dip temperatures of $g_{CG}$ can also be used, and the data are plotted in the right panel of \Rfig{Tc}. The crossing temperature of $g_{CG}$ obeys the scaling form \Req{T_cross} with $T_{c}=T_{CG}$ and $\theta=\theta_{CG}=\omega_{CG}+1/\nu_{CG}$, whereas the dip temperature is expected to follow the scaling form,
\be
T_{dip}(L)=T_{CG}+c_{1}L^{-\frac{1}{\nu_{CG}}}+c_{2}L^{-\theta_{CG}}.
\Leq{T_dip}
\ee
Conventionally, the sub-leading correction-to-scaling term, $c_{2}L^{-\theta_{CG}}$, is dropped because it gives a relatively smaller contribution than that of the leading correction-to-scaling term $c_{1}L^{-1/\nu_{CG}}$ for larger $L$. In the range of system sizes of our simulations, however, we need to take into account this correction term for describing the size dependence of $T_{cross}$, since the leading term $c_{1}L^{-1/\nu_{CG}}$ should describe the increase of the dip temperature with the system size, which is not observed in our simulation. (Such an increase of $T_{dip}(L)$ was indeed observed in a recent simulation of the 4D Heisenberg SG~\cite{Kawamura:12}.) The behavior originates from the fact that both the dip and the crossing temperatures of $g_{CG}$ should converge to a common value, $T_{CG}$, each with an exponent $1/\nu$ and $\theta (> 1/\nu)$, while the crossing temperature always lies above the dip one.

The combined fit of the crossing and the dip temperatures of $g_{CG}$ based on \Reqs{T_cross}{T_dip} with a common $T_{CG}$ and a common $\theta_{CG}$ 
using the series $s=2,3/2,4/3$ and $5/3$ yields $T_{CG}=0.308\pm 0.005$ and $\theta_{CG}=0.88 \pm 0.03$. More sensible error bars are obtained by only using the $s=2$ series, to find $T_{CG}=0.313^{+0.013}_{-0.018}$ and $\theta_{CG}=0.91 \pm 0.10$. Those error bars are again calculated from the total $\chi$-square value against the assumed $T_{CG}$ value, and the result is given in the main panel of \Rfig{chi2-both}. Note that the leading term in \Req{T_dip} tends to be masked by the correction term, {\it i.e.\/}, $c_1$ tends to be considerably smaller than $c_2$.

%
%

The SG and the CG transition temperatures $T_{SG}=0.275^{+0.013}_{-0.052}$ and $T_{CG}=0.313^{+0.013}_{-0.018}$ estimated here are well consistent with the values obtained from the order parameter given in \Rfig{L-q}. Our error analysis indicates that $T_{SG}$ and $T_{CG}$ are different by more than two $\sigma$s, and are indeed separate. These observations certainly speak for the occurrence of the spin-chirality decoupling in the 3D {\it XY\/} SG. The difference in the CG and the SG transition temperatures is about $10$ percent, which is comparable with the corresponding value of the 3D Heisenberg SG~\cite{Viet:09}. 

In \Rfig{gziCGoverSG}, we plot the ratio of the CG and the SG correlation lengths, $\xi_{CG}/\xi_{SG}$. 
\begin{figure}[htbp]
\includegraphics[width=1\columnwidth,height=0.8\columnwidth]{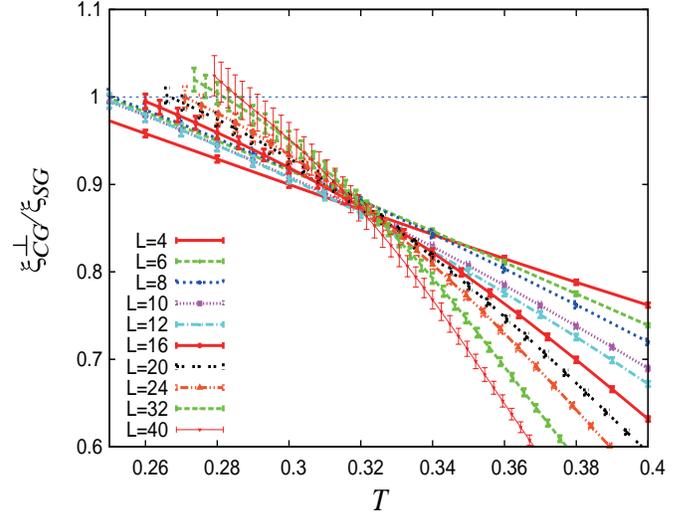}
\caption{\Lfig{gziCGoverSG}(Color online) The temperature and size dependence of the ratio of the CG and the SG correlation lengths $\xi^{\perp}_{CG}/\xi_{SG}$. 
}
\end{figure}
For smaller sizes of $6\leq L \leq 12$, the ratio curves  are almost size-independent~\cite{Lee:03} at lower temperatures, while for intermediate sizes of $16\leq L \leq 24$ the curves start to splay out but the tendency is still small~\cite{Pixley:08}. For larger sizes of $L=32$ and $40$, the tendency becomes stronger and the ratio exceeds unity at low temperatures, which supports the spin-chirality decoupling ansatz. The intersection point of the ratio curves between different sizes lie around $T\sim 0.31$, which is consistent with our estimate of $T_{CG}=0.313^{+0.013}_{-0.018}$.

Now, to get further insight into the cause of the pathological behavior we encountered in $\xi_{CG}$, we discuss the CG and the SG spatial correlations. The SG and the CG spatial correlation functions, $C_{s}(x)$ and $C_{\kappa}(x)$, are defined by
\be
&&
C_{s}(|\V{r}_{i}-\V{r}_{j}|)=\sum_{\alpha,\beta}
\lsb
\Ave{S^{(1)}_{i\alpha}S^{(2)}_{i\beta}S^{(1)}_{j\alpha}S^{(2)}_{j\beta}}
\rsb,
\\
&&
C_{\kappa}(|\V{r}_{p}-\V{r}_{q}|)=
\lsb
\Ave{\kappa^{(1)}_{p \perp \mu}\kappa^{(2)}_{p \perp \mu}\kappa^{(1)}_{q \perp \mu}\kappa^{(2)}_{q\perp \mu}} 
\rsb.
\ee
The computed $C_{s}$ and $C_{\kappa}$ for $L=16$ lattices under periodic boundary conditions (BCs) are shown in \Rfig{corfunc}  on a semi-logarithmic scale for several temperatures. Note that the CG correlation function $C_{\kappa}$ is normalized by its local amplitude so as to give unity at $x=0$. The leveling-off of the data at larger $x$ is a finite-size effect due to the imposed periodic BC.
\begin{figure}[htbp]
\includegraphics[width=1\columnwidth,height=0.8\columnwidth]{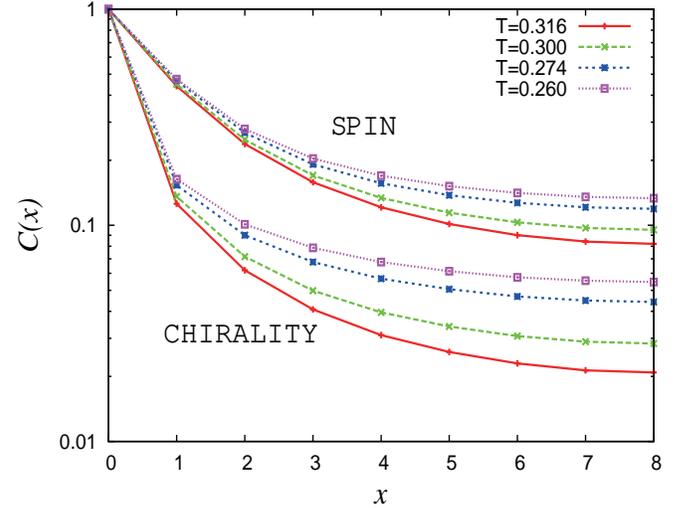}
\caption{\Lfig{corfunc}(Color online) The spatial correlation functions of the SG and of the CG overlaps for several temperatures around $T_{CG}\simeq 0.313$ and $T_{SG}\simeq 0.275$. The upper curves are for the spin overlap and the lower ones are for the chiral one. The lattice is $L=16$ under periodic BC. Total number of samples is $N_{s}=500$. Error bars are omitted. Note the logarithmic scale of the ordinate.
}
\end{figure}

 An important point is to be noticed here is the large difference between $C_{s}(x)$ and $C_{\kappa}(x)$ in their magnitudes. Namely, $C_{\kappa}(x)$ drops rapidly from unity in the small-$x$ region of a few lattice spacings even below $T_{CG}$, and becomes by an order of magnitude smaller than $C_{s}(x)$ for larger $x$. Such a sharp drop of the spatial correlations at short length is much less pronounced in $C_{s}(x)$. This feature of the CG correlations might cause some problems in defining the finite-size correlation length  $\xi_{CG}$ based on \Req{xi_CG}, at least for small lattices treated in our present simulation. This is because this definition of $\xi_{CG}$ contains the $k=0$ part $[\Ave{q_{\kappa}^{\mu}(0)^2}]$, which is essentially an equal-weight sum of CG correlation functions, $\int dx C_{\kappa}(x)$. Since $C_{\kappa}(x)$ in the large-$x$ region, which should govern the true CG correlation length, is much smaller in magnitude than that in the small-$x$ region, the latter contribution not playing an essential role in the CG correlation length might make a major contribution to $[\Ave{q_{\kappa}^{\mu}(0)^2}]$, and mask or obscure the asymptotic behavior of the CG correlation length. Thus, the standard definition of the finite-size correlation length, \Req{xi_CG}, may be inappropriate to treat the CG correlation length, when the system size is small. A similar observation was also made in the 4D Ising SG in a magnetic field~\cite{Leuzzi:09}. To overcome this difficulty, another dimensionless quantity was already proposed~\cite{Leuzzi:09}, but the examination of the quantity is beyond the scope of the present paper.

Next, we turn to the quantities which probe the phase-space structure of the ordered state, including the overlap distribution and the non-self-averageness parameter. In \Rfig{P}, we show the distributions of the spin and the chiral overlaps at a temperature $T=0.2792$, which lies below $T_{CG}$ and very close to (slightly above) $T_{SG}$.
\begin{figure*}[htbp]
\includegraphics[width=0.7\columnwidth,height=1\columnwidth,angle=-90]{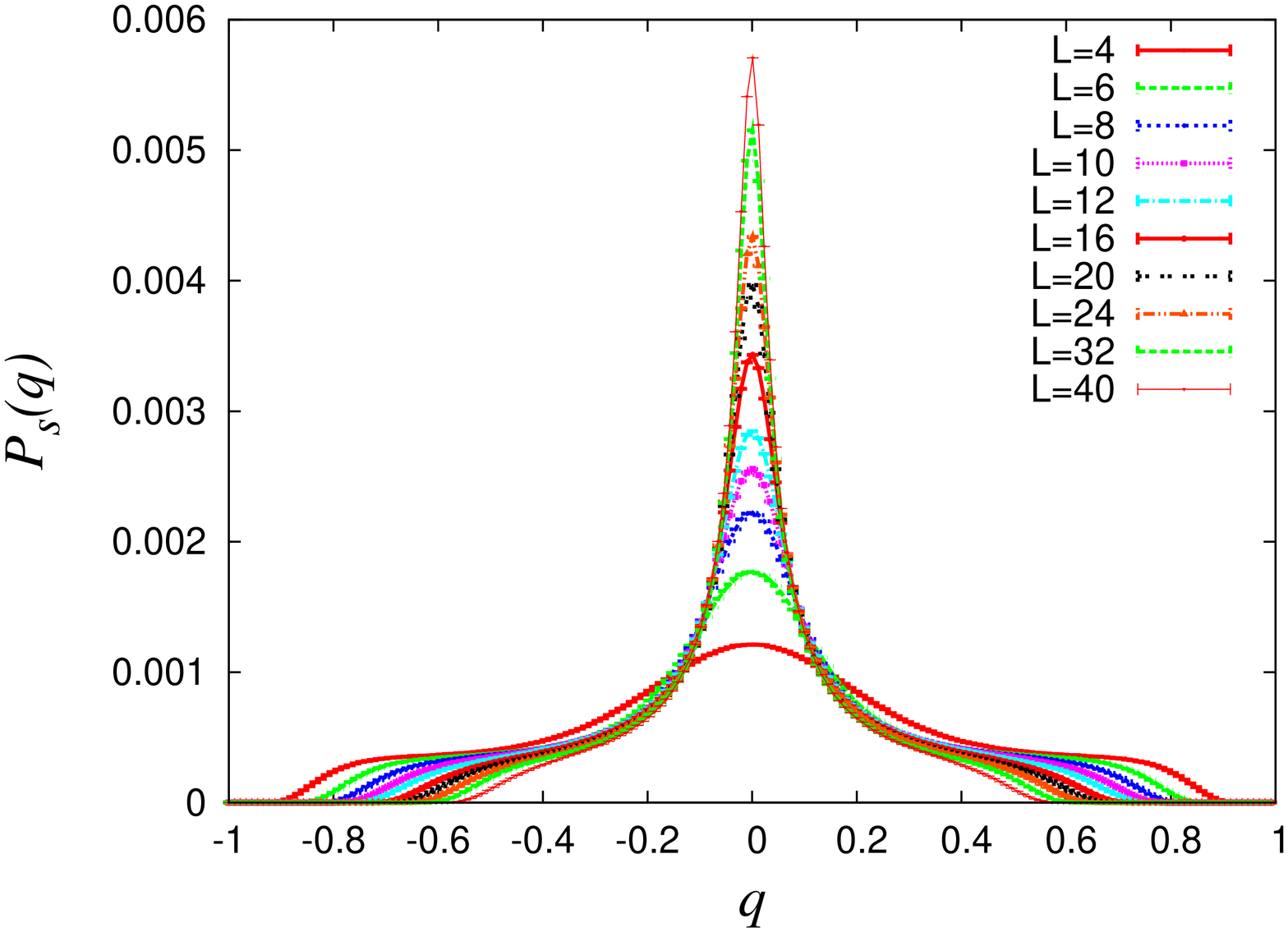}
\includegraphics[width=0.7\columnwidth,height=1\columnwidth,angle=-90]{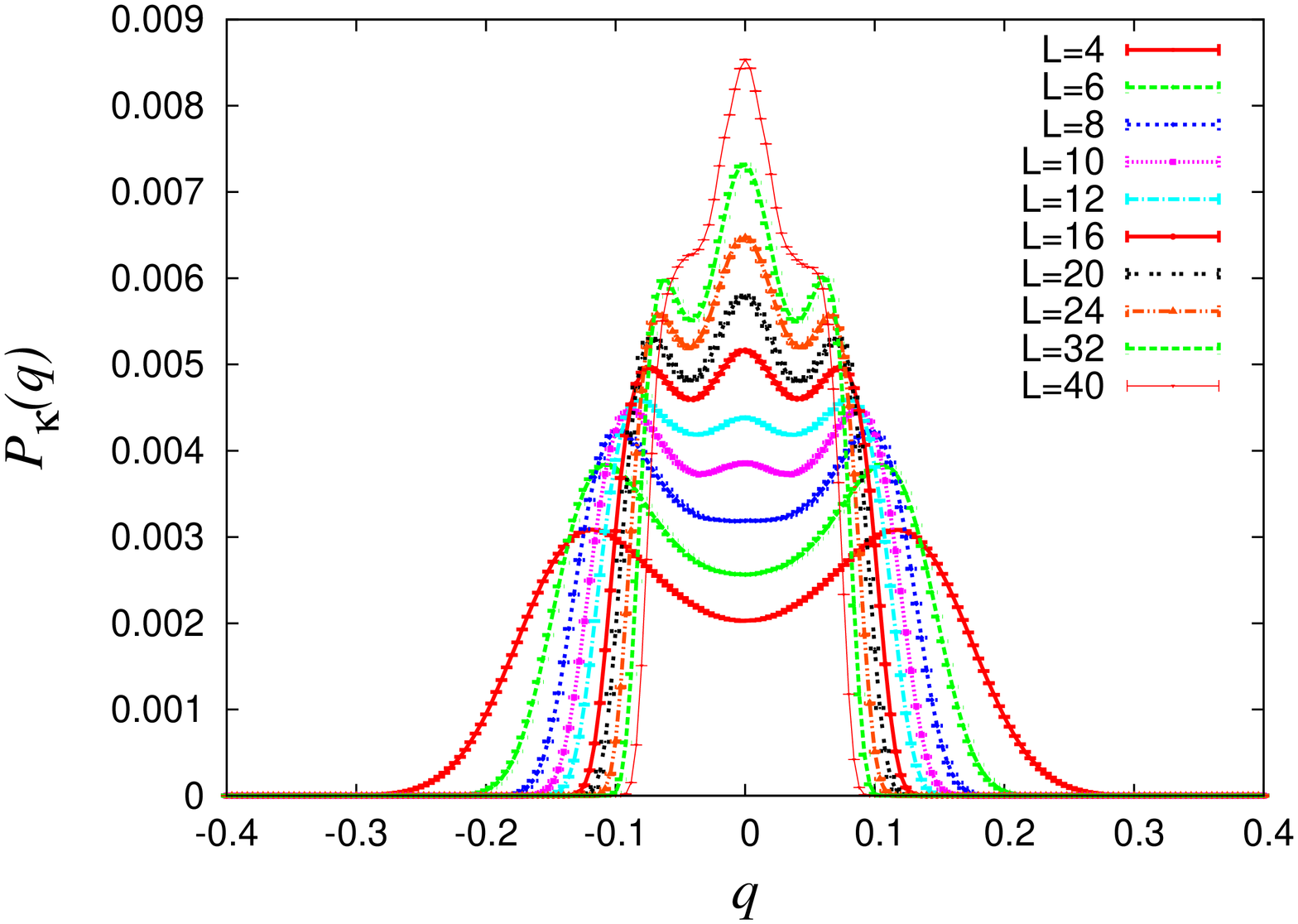}
\caption{\Lfig{P}(Color online) The spin diagonal overlap distribution (left), and the chiral overlap distribution (right), at a temperature $T=0.2792$ which lies in the CG ordered state, {\it i.e.\/} below $T_{CG}\simeq 0.308$ but slightly above $T_{SG}\simeq 0.274$. A typical 1RSB behavior is observed in the chiral overlap distribution.
}
\end{figure*}
For $L \geq 10$, the chiral-overlap distribution $P_{\kappa}$ shown in the right panel of \Rfig{P} exhibits a central peak in addition to the side peaks corresponding to the CG EA order parameter $\pm q_{CG}^{EA}$.  As the system size $L$ increases, all the peaks grow in their height and become narrower in their width. This implies that all these peaks would remain in the thermodynamic limit. These features are nothing but the character of the 1RSB, and are consistent with the occurrence of a negative dip in $g_{CG}$. Similar behaviors were observed in several types of Heisenberg SG before~\cite{Imagawa:03,Viet:09}, but the side peaks observed in our present simulation for the {\it XY} SG seem sharper than those observed in the Heisenberg SG.

 By contrast, the spin-overlap distribution shown in the left panel of \Rfig{P} exhibits a shoulder-like structure only for small sizes, which tends to be suppressed as the system size increases. As the growing side peaks located at $\pm q_{SG}^{EA}$ are expected in the SG ordered phase~\cite{Kawamura:01-1}, this observation is consistent with the absence of the SG order at this temperature $T=0.2792$, which is indeed compatible with our estimate above, $T_{SG}=0.275^{+0.013}_{-0.052}$.

 In \Rfig{A}, the SG and CG non-self-averagingness $A$ parameters are plotted  against the temperature for various system sizes.
\begin{figure*}[htbp]
\includegraphics[width=0.7\columnwidth,height=1\columnwidth,angle=-90]{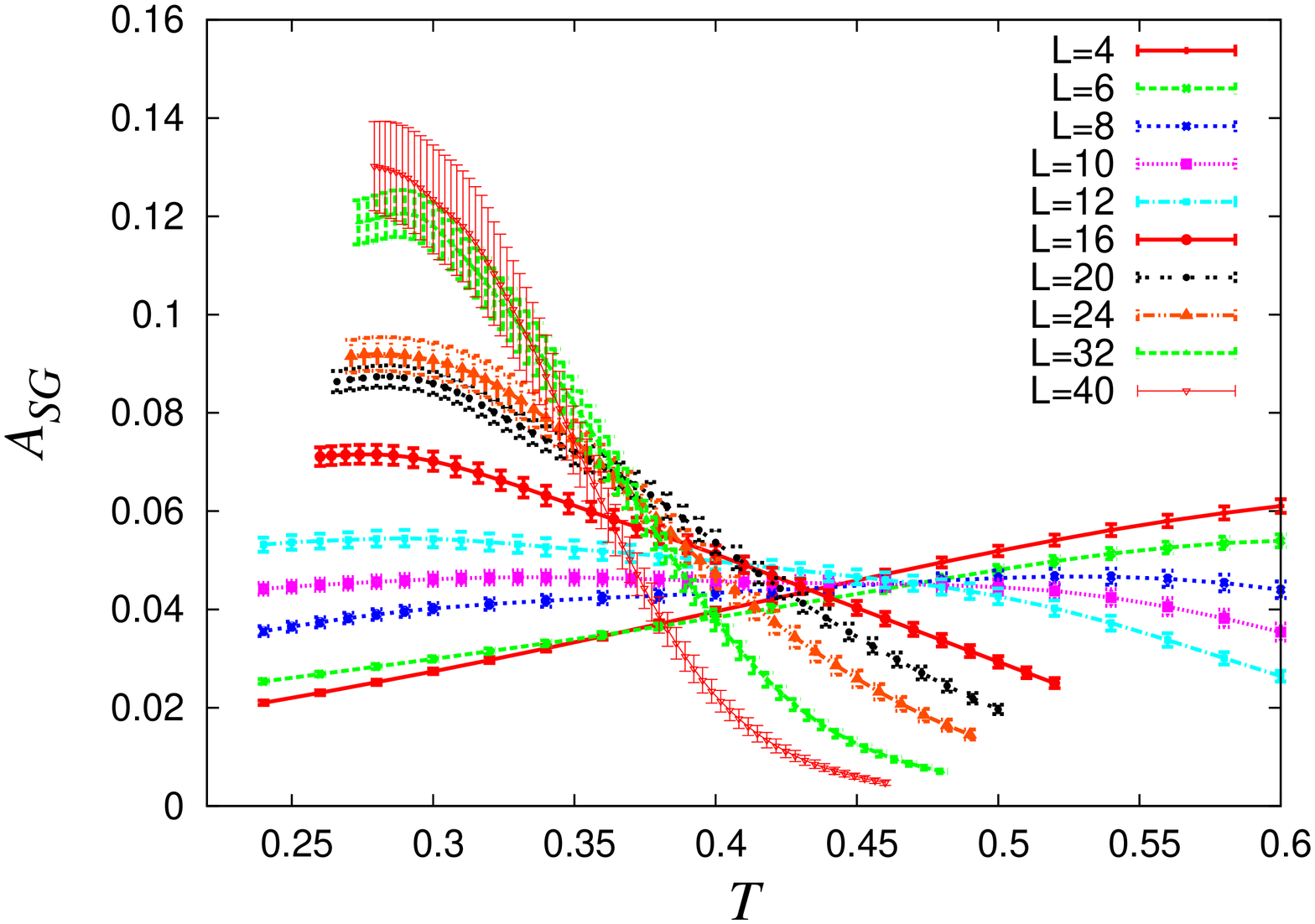}
\includegraphics[width=0.7\columnwidth,height=1\columnwidth,angle=-90]{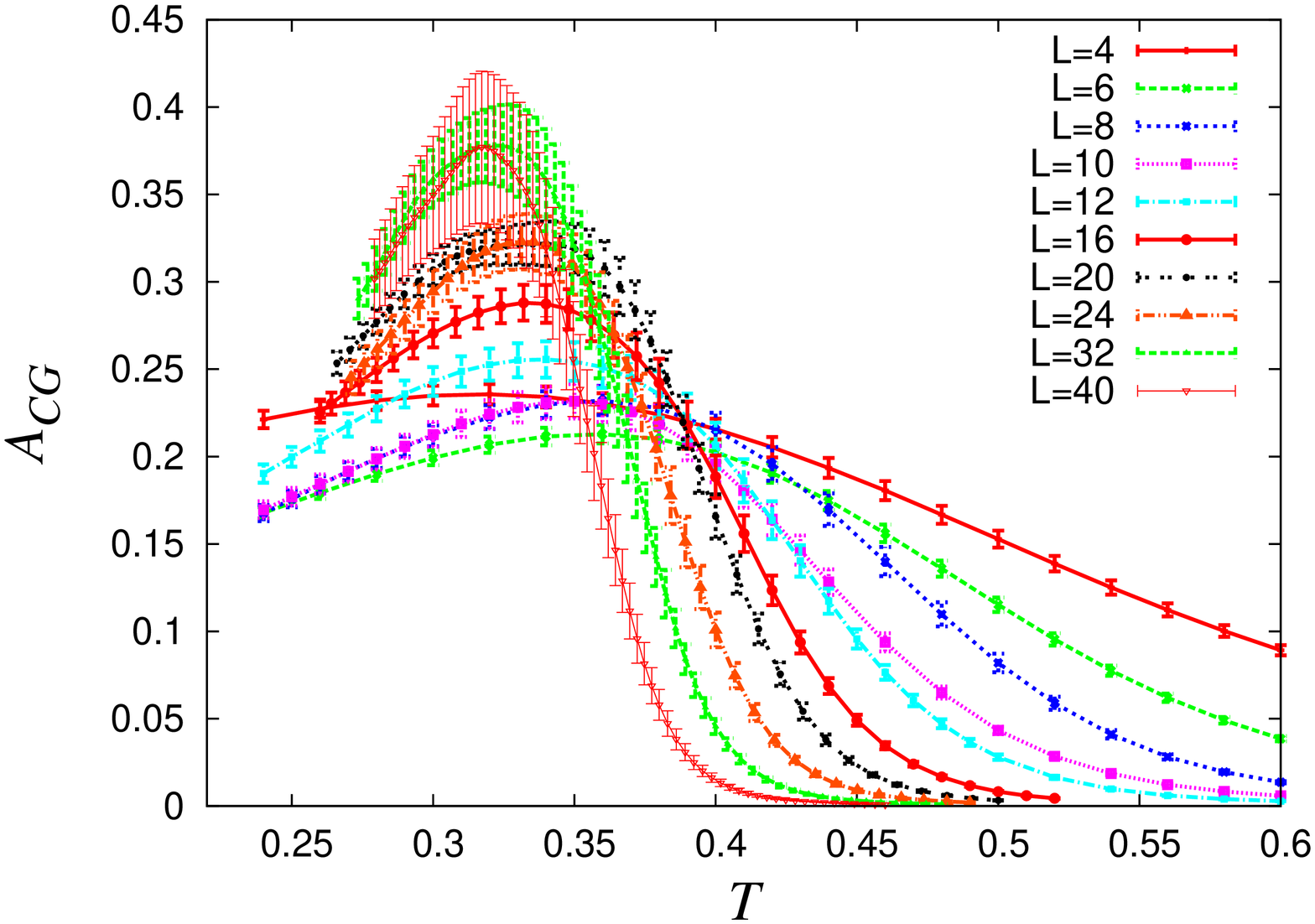}
\caption{\Lfig{A}(Color online) The temperature and size dependence of the non-self-averageness $A$ parameters of the SG (left), and of the CG (right).
}
\end{figure*}
As can be seen from the figure, $A_{CG}$ of different sizes intersect around $T\sim T_{CG}$. A prominent peak is observed on the lower temperature side, which grows as the system size increases. This suggests that the self-averageness of the system is broken below $T_{CG}$ and that the CG ordered phase is non-self-averaging. This is quite consistent with the 1RSB nature of the CG phase, as already signaled by the central peak in $P_{\kappa}$ and by the negative dip of $g_{CG}$. The parameter $A_{SG}$ also exhibits an intersection around $T=T_{CG}$, even though the SG order is still absent at $T=T_{CG}$. As already noted in \Rsec{quantity}, this is not surprising since a finite $A_{SG}$ just means the non-self-averageness of $\chi_{SG}$ which originates from the CG transition involving the phase-space narrowing associated with the 1RSB. Hence, the intersection occurring around $T_{CG}$ is completely compatible with the spin-chirality decoupling. It might be interesting to point out that,  for larger $L$, the peak of the $A$ parameter seems to be located near the respective transition temperature, {\it i.e.\/}, near $T_{CG}$ for $A_{CG}$ while near $T_{SG}$ for $A_{SG}$. Such a difference between the behaviors of $A_{CG}$ and $A_{SG}$ might be consistent with the spin-chirality decoupling.

 We also show the SG and CG $G$ parameters in \Rfig{G}. In case of the Ising SG, the $G$ parameter of large enough lattices exhibits a crossing at the transition temperature and tends to $1/3$ in the $T\rightarrow 0$ limit~\cite{Marinari:98}. The behavior observed here resembles such a behavior of the Ising SG. 
\begin{figure*}[htbp]
\includegraphics[width=0.7\columnwidth,height=1\columnwidth,angle=-90]{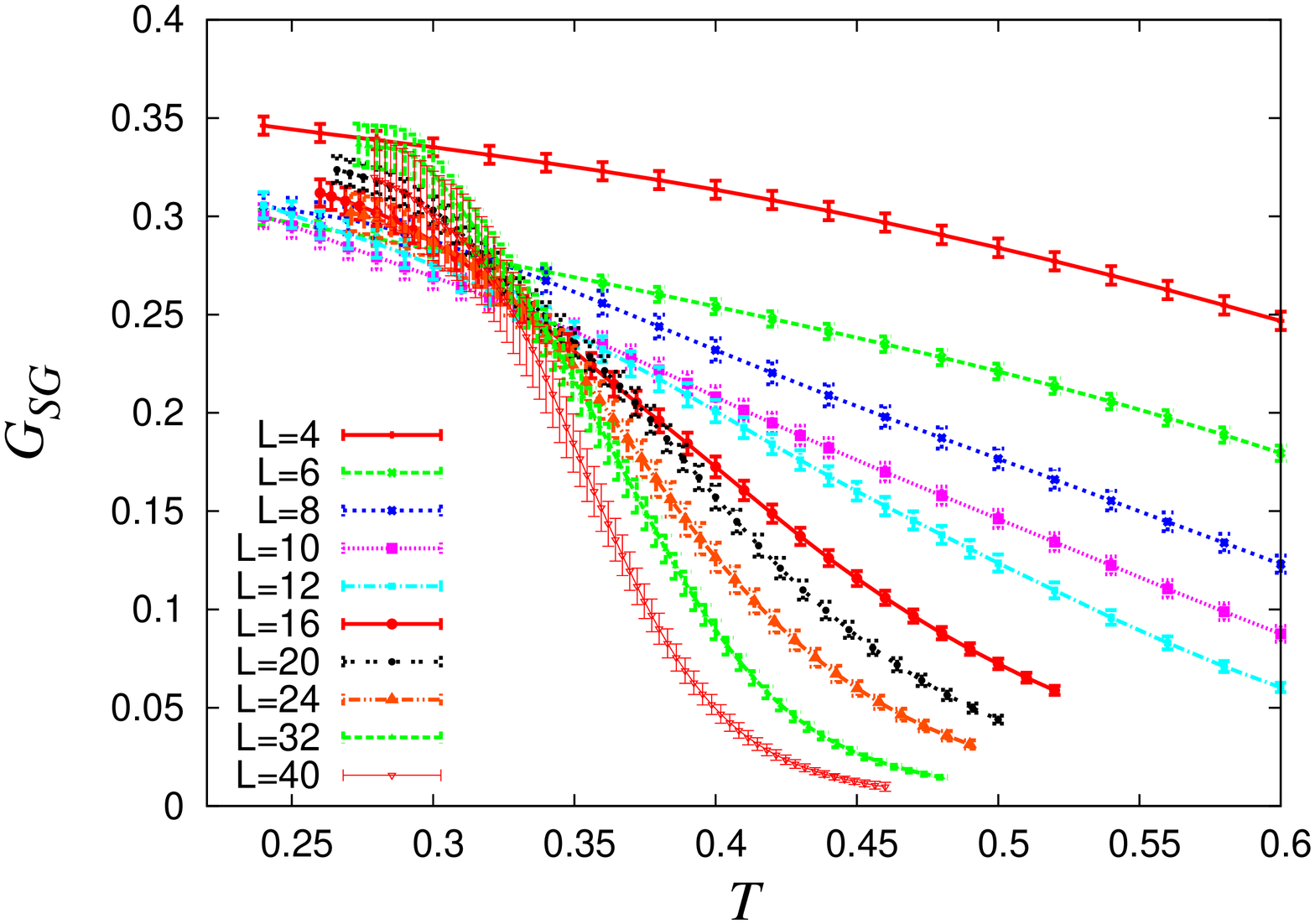}
\includegraphics[width=0.7\columnwidth,height=1\columnwidth,angle=-90]{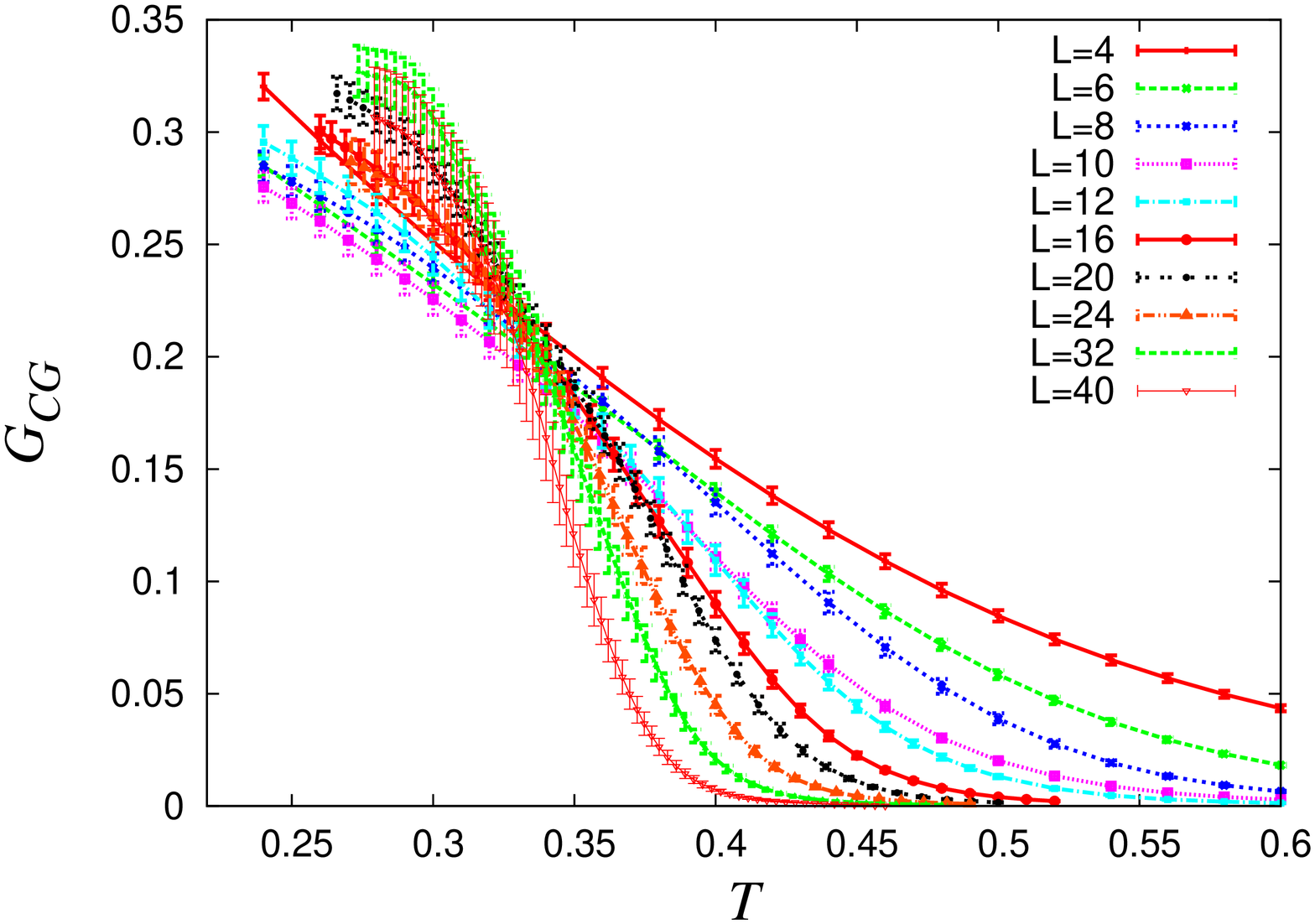}
\caption{\Lfig{G}(Color online) The temperature and size dependence of the $G$ parameters of the SG (left), and of the CG (right).
}
\end{figure*}
Note that, although the intersections of these $A$ and $G$ parameters can also be used in estimating the transition temperature in principle, the data tend to be noisy and not suited to precisely locate $T_{CG}$. The same suggestion was made for the 3D Ising SG~\cite{Ballesteros:00,Palassini:03} and for the 3D Heisenberg SG~\cite{Viet:09}.


\section{CRITICAL PROPERTIES}

\Lsec{critical}

In this section, we study the critical properties of the SG and the CG transitions based on the finite-size scaling analysis. To estimate the CG critical exponents, we use both the CG Binder parameter and the CG susceptibility. On the other hand, we use the SG susceptibility only to estimate the SG critical exponents, the reason of which will be explained below. 

Let us start from the CG criticality. The standard finite-size scaling forms of the CG Binder parameter $g_{CG}$ and of the CG susceptibility $\chi_{CG}$ are given by
\be
g_{CG}=\T{X}\lb \lb T-T_{CG}\rb L^{1/\nu_{CG}} \rb,
\Leq{gCG-FSS}
\\
\chi_{CG}=L^{2-\eta_{CG}}\T{Y}\lb 
\lb T-T_{CG}\rb L^{1/\nu_{CG}}
\Leq{chiCG-FSS}
\rb,
\ee
where $\T{X}$ and $\T{Y}$ are appropriate scaling functions.

For the CG susceptibility $\chi_{CG}$,  a good data collapse can be obtained by the two-parameter fits based on \Req{chiCG-FSS}, with $\nu_{CG}=1.28$ and $\eta_{CG}=0.34$. Note that these values are obtained via the Baysian scaling analysis (BSA), which enables us to estimate the critical exponents in an unbiased way~\cite{Harada:11}, after fixing the transition temperature to the value obtained in the previous section $T_{CG}=0.313$. By changing the assumed value of $T_{CG}$ in the range of the associated error bar, we obtain the error bars of $\nu_{CG}$ and $\eta_{CG}$ as $\nu_{CG}=1.28^{+0.12}_{-0.03}$ and $\eta_{CG}=0.34^{+0.40}_{-0.34}$.

 The crossing temperature of the CG Binder parameter $g_{CG}$  still exhibits appreciable size dependence, indicating the necessity to invoke the correlation-to-scaling term in the finite-size scaling. With the correction term, the scaling form of $g_{CG}$ is modified as
\be
g_{CG}=\T{X}\lb \lb T-T_{CG}\rb L^{1/\nu_{CG}} \rb
\lb 1+aL^{-\omega_{CG}}\rb,
\ee
where $a$ is a numerical constant. The BSA analysis based on this form yields $\nu_{CG}=1.36^{+0.15}_{-0.37}$ and $\omega_{CG}=0.48^{+0.17}_{-0.14}$ with a constant $a=10^{+90}_{-7}$, which is consistent with $\nu_{CG}=1.28^{+0.12}_{-0.03}$ estimated above from the CG susceptibility. The value of the nonuniversal constant $a$  tends to be pretty large, though it contains rather large error bar.

These values lead to $(1/\nu_{CG})+\omega_{CG}\approx 1.2\pm 0.4$, which is consistent with $\theta_{CG}=0.91\pm 0.10$ obtained in the previous section. The resultant finite-size scaling plot is given in the left panel of \Rfig{CG-FSS}. 
\begin{figure*}[htbp]
\includegraphics[width=1\columnwidth,height=0.75\columnwidth]{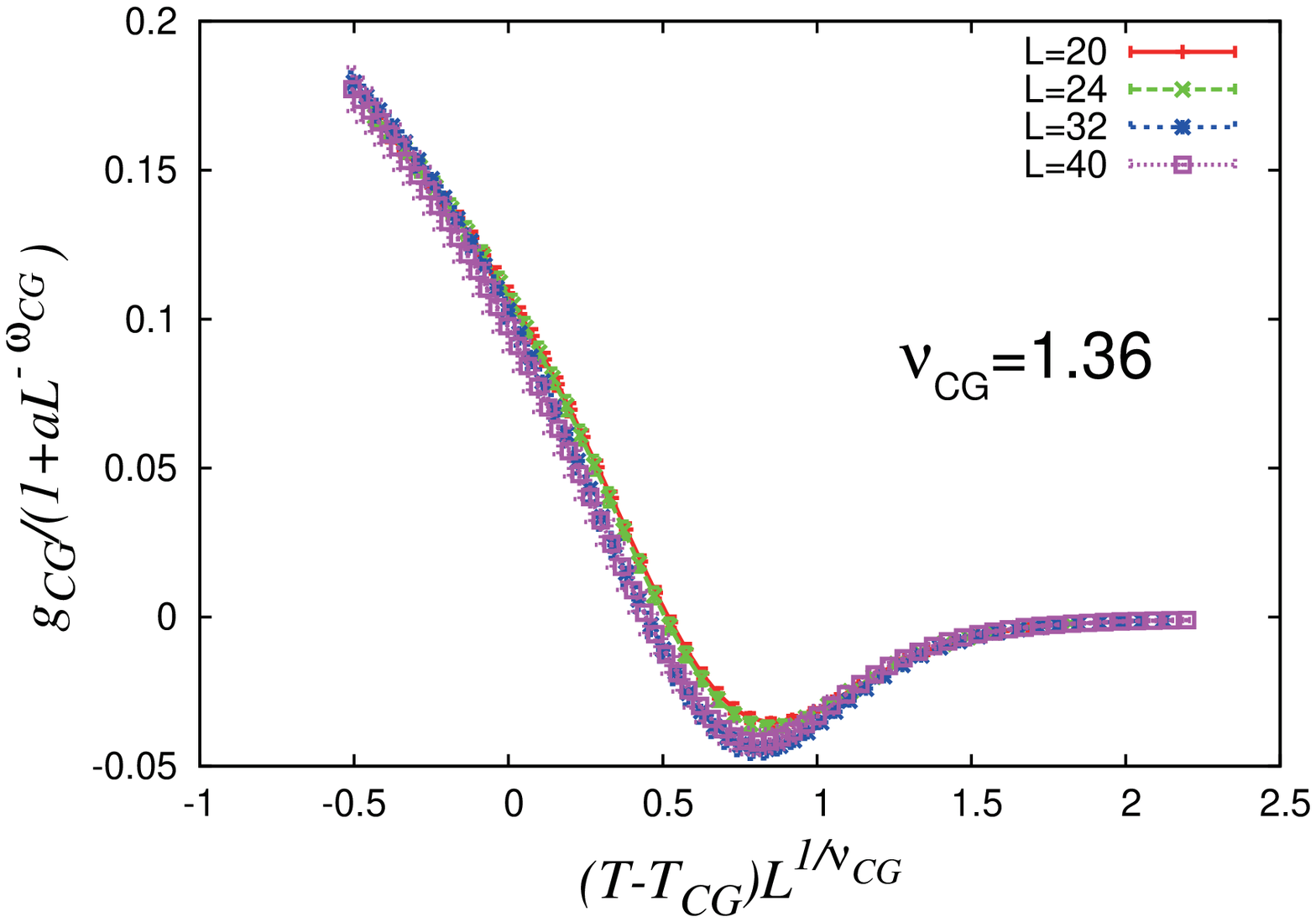}
\includegraphics[width=1\columnwidth,height=0.75\columnwidth]{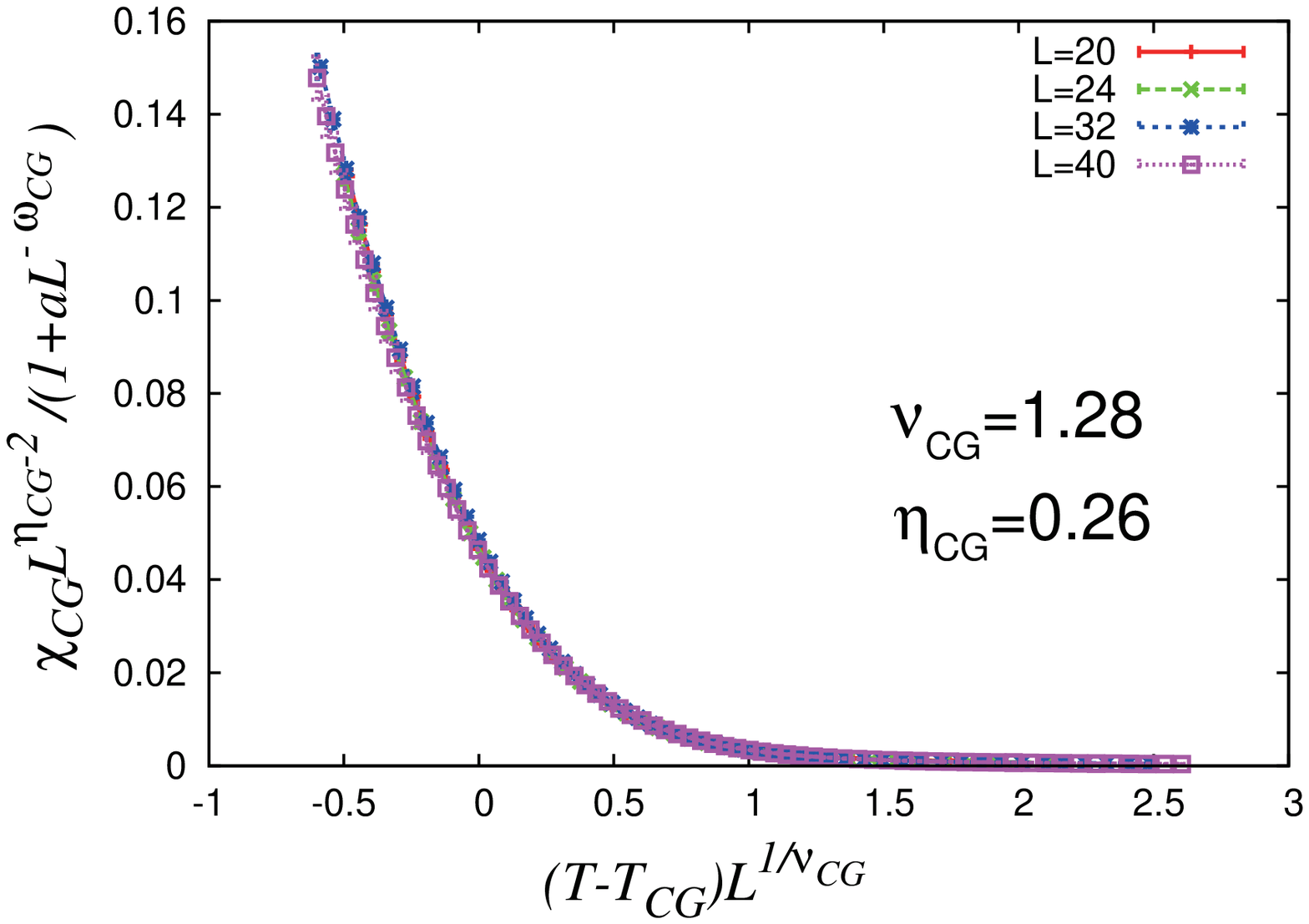}
\caption{\Lfig{CG-FSS}(Color online) Finite-size-scaling plots of the CG Binder parameter (left), and of the CG susceptibility (right), where the correction-to-scaling effect is taken into account. The CG transition temperature is fixed to $T_{CG}=0.313$. Best fit is obtained for $\nu_{CG}=1.36$ and $\omega_{CG}=0.48$ with $a=10^{+90}_{-7}$ for $g_{CG}$ (left), and $\nu_{CG}=1.28$, $\eta_{CG}=0.26$ and $\omega_{CG}=0.3$ with $a=0.5^{+0.5}_{-0.1}$ for $\chi_{CG}$ (right).
}
\end{figure*}

 We examine the scaling form with the correction term also for the CG susceptibility,
\be
\hspace{-7mm}\chi_{CG}=L^{2-\eta_{CG}}\T{Y}\lb 
\lb T-T_{CG}\rb L^{1/\nu_{CG}}\rb
\lb 1+a'L^{-\omega_{CG}}\rb.
\ee
Based on this form, we get $\nu_{CG}=1.28^{+0.12}_{-0.03}$, $\eta_{CG}=0.26^{+0.29}_{-0.26}$, and $\omega_{CG}=0.32^{+0.21}_{-0.10}$ with a nonuniversal constant $a'=0.5^{+0.5}_{-0.1}$. These values are consistent with the values obtained above without invoking the correction term. The resultant finite-size scaling plot is given in the right panel of \Rfig{CG-FSS}. 

Next, we move to the SG criticality. As in the CG case, the standard scaling form of the SG susceptibility is given by
\be
\chi_{SG}=L^{2-\eta_{SG}}\T{Y}\lb 
\lb T-T_{SG}\rb L^{1/\nu_{SG}}
\Leq{chiSG-FSS}
\rb.
\ee
A good data collapse of the SG susceptibility is obtained based on this form, with the resultant exponents  $\nu_{SG}=1.23^{+0.28}_{-0.07}$ and $\eta_{SG}=-0.42^{+0.13}_{-0.40}$. We also examine the effect of the correction-to-scaling based on the form,
\be
\hspace{-7mm}\chi_{SG}=L^{2-\eta_{SG}}\T{Y}\lb 
\lb T-T_{SG}\rb L^{1/\nu_{SG}}\rb
\lb 1+a''L^{-\omega_{SG}}\rb,
\ee
to get $\nu_{SG}=1.22^{+0.26}_{-0.06}$ and $\eta_{SG}=-0.54^{+0.24}_{-0.52}$, and $\omega_{SG}=0.89^{+0.11}_{-0.05}$ with a nonuniversal constant $a''=3^{+6}_{-2}$. The associated scaling plot at $T=T_{SG}=0.275$ is given in \Rfig{SG-FSS}. 
\begin{figure}[htbp]
\includegraphics[width=1\columnwidth,height=0.75\columnwidth]{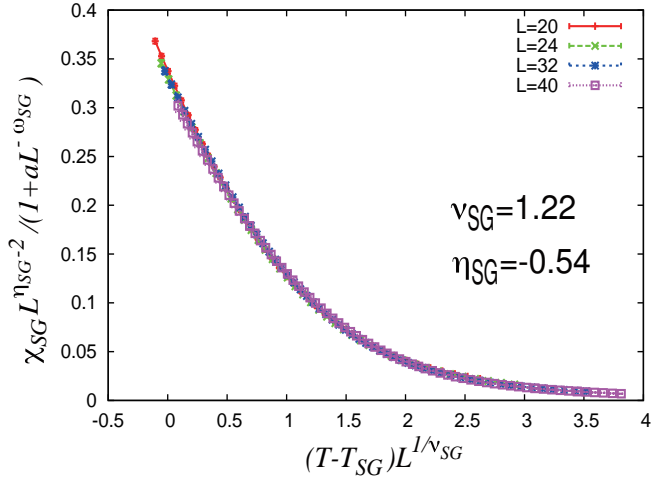}
\caption{\Lfig{SG-FSS}(Color online) Finite-size-scaling plot of the SG susceptibility with the correction-to-scaling term.  The SG transition temperature is fixed to $T_{SG}=0.275$. Best fit is obtained for $\nu_{SG}=1.22$, $\eta_{SG}=-0.54$, and $\omega_{SG}=0.89$.
}
\end{figure}
 We also examined the SG correlation-length ratio but ended up with an unphysical result of negative $\omega_{SG}$, so we do not quote it here. This inadequacy may partly be due to the fact that the estimated $T_{SG}$ is located out of the range of the simulated temperature range.

Summarizing the above results, we finally quote as our best estimates of the CG exponents as
\be
\nu_{CG}=1.36^{+0.15}_{-0.37},\,\, \eta_{CG}=0.26^{+0.29}_{-0.26},
\ee
and the SG exponents as
\be
\nu_{SG}=1.22^{+0.26}_{-0.06},\,\, \eta_{SG}=-0.54^{+0.24}_{-0.52}.
\ee

The estimated CG critical exponents are compatible with the ones reported in earlier literature on the 3D {\it XY\/} SG~\cite{Kawamura:95-1,Kawamura:01-1,Granato:04,Nakamura:06,Pixley:08}. They are also quite close to the values of the 3D Heisenberg SG~\cite{Viet:09}, whereas they are largely different from the values of the 3D Ising SG, $\nu=2.5 \sim 2.7$ and $\eta=-0.38 \sim -0.40$~\cite{Campbell:06,Hasenbusch:08} in spite of a common $Z_{2}$ symmetry between  the Ising spin and the chirality of the present model. The types of the RSB  in the Ising SG and  that in the {\it XY\/} or the Heisenberg SG may explain this difference, {\it i.e.\/}, the full RSB in the former versus the 1RSB in the latter. The phase-space structure of the {\it XY} and the Heisenberg SGs is essentially different from that of the Ising SG. Such speculation also leads to another question: what causes the difference in the RSB types among the Ising, the {\it XY\/} and the Heisenberg SGs? A possible explanation might be that the chirality-chirality interaction has a long-range nature different from the Ising one. Further study is needed to clarify these points.


\section{Summary and Discussion}

In this paper, we studied equilibrium ordering properties of the 3D isotropic {\it XY} SG by means of extensive MC simulations, up to the linear size $L=40$. Examining various physical quantities including the glass order parameter, the Binder parameter, the correlation-length ratio, the overlap-distribution function and the non-self-averageness parameter, we succeeded in giving reasonable numerical evidence that the SG and the CG transitions occur at two different temperatures. The SG and CG critical temperatures estimated from the correlation-length ratio and the Binder parameter are $T_{SG}=0.275^{+0.013}_{-0.052}$ and $T_{CG}=0.313^{+0.013}_{-0.018}$, respectively. Since the difference is more than two $\sigma$s, the difference is likely to be statistically relevant. Furthermore, our independent estimate of the CG and the SG transition temperatures based on the glass order parameter $q^{(2)}$ turned out to be entirely consistent with the estimates from the correlation-length ratio and the Binder parameter. The difference between $T_{CG}$ and $T_{SG}$ is about $10$ percent, which is comparable with the difference observed in the 3D Heisenberg SG.

 Our conclusion of the occurrence of the spin-chirality decoupling in the model is in apparent contrast to that of the recent simulation by Pixley and Young~\cite{Pixley:08}.  The main cause of the difference is that their analysis was mainly based on the correlation-length ratios. We have also confirmed that, as observed by Pixley and Young, the crossing points of the CG correlation-length ratio behave in not much different manner from the SG ones, which are seemingly consistent with a simultaneous SG and CG transition. Our present quantitative analysis, however, revealed that the extrapolated $T_{cross}(L)$ of the $\xi_{CG}/L$ lead to an unphysical estimate of $T_{CG}$, {\it i.e.\/}, a negative one, whereas that of $\xi_{SG}/L$ lead to a reasonable estimate of $T_{SG}$. It turned out that such a behavior of the CG correlation-length ratio is inconsistent with that of other quantities such as the CG order parameter and the CG Binder parameter. In particular, the size dependence of the glass order parameter speaks for the successive CG and SG transitions occurring at two different temperatures, as seen in \Rfig{L-q}. Our attitude is that the order parameter among various quantities is expected to give a stable and reliable result, since it is the most fundamental quantity in describing phase transition.  We then argued that the observed ill behavior of the finite-size CG correlation length may originate from the finite-size effect associated with a significant short-length drop-off of the spatial CG correlations. Pixley and Yound also suggested a possibility that the 3D {\it XY\/} SG is marginal, {\it i.e.\/}, the lower critical dimension is close to three. Our data, especially the glass order parameter shown in \Rfig{L-q} and the critical scaling shown in \Rfigss{CG-FSS}{SG-FSS}, reasonably rules out such a possibility.

 The critical properties of the SG and the CG orderings were also examined by means of the finite-size-scaling analysis. By controlling the correction-to-scaling effect, we obtained the CG critical exponents as $\nu_{CG}=1.36^{+0.15}_{-0.37}$ and $\eta_{CG}=0.26^{+0.29}_{-0.26}$. These values are close to the corresponding Heisenberg SG values, but are quite different from the Ising SG values in spite of a common $Z_2$ symmetry. The SG critical exponents were also estimated to be $\nu_{SG}=1.22^{+0.26}_{-0.06}$ and $\eta_{SG}=-0.54^{+0.24}_{-0.52}$, which were consistent with the earlier estimates. 

The RSB nature of the ordered state was probed via the Binder parameter, the overlap distribution and the non-self-averageness parameter. All the quantities consistently point to the 1RSB in the model. Physical significance of the 1RSB feature was already discussed in the Heisenberg case from several perspectives~\cite{Kawamura:10}. It would also be interesting to examine the possible 1RSB properties in real materials related to the {\it XY} SG such as granular cuprate superconductors~\cite{Deguchi:09,Matsuura:95,Yamao:99,Papadopoulou:99,Hagiwara:05}, together with the successive transitions and the associated critical properties.

 As mentioned, the CG transition of the {\it XY\/} SG model and the Ising SG transition belong to different universality classes in spite of a common $Z_2$ symmetry between the chirality and the Ising spin variable. We speculate that this might originate from the difference in the type of the RSB in the two systems. The 3D Ising SG is believed to exhibit a full RSB~\cite{Bhatt:85,Marinari:98,Marinari:99}, though some counter opinions also exist~\cite{Fisher:86,Bray:87,Drossel:00,Jorg:08}. By contrast, the 3D {\it XY} SG exhibits the 1RSB. These different types of RSB may be related to the difference in the critical properties of  the Ising SG transitions and the CG transition of the {\it XY\/} SG. Furthermore, since such a 1RSB-like feature is also observed in the 3D Heisenberg SG~\cite{Viet:09}, the above consideration suggests a possibility that the CG transitions of the {\it XY\/} and the Heisenberg SGs are actually the same, which are indeed consistent with our present numerical results.


\begin{acknowledgments}
The authors are grateful to T. Okubo, G. Parisi and H. Yoshino for valuable discussions and comments. Especially, the discussion on the asymptotic form of the SG Binder parameter given in appendix owes much to T. Okubo. G. Parisi also gave useful suggestions concerning the property of the CG spatial correlations argued in \Rsec{result}. This study is supported by Grant-in-Aid Scientific Research on Priority Areas `Novel State of Matter Induced by Frustration' (No. 19052006 and No. 19052008). We thank ISSP, University of Tokyo, and YITP, Kyoto University for providing us with the CPU time.
\end{acknowledgments}


\appendix
\section{The behavior of the spin-glass Binder parameter in the thermodynamic limit}

In this appendix, we examine how the SG Binder parameter $g_{SG}$ in the thermodynamic limit behaves across $T_{CG}$ and $T_{SG}$ in the occurrence of the spin-chirality decoupling.

 In the CG phase realized at $T_{SG}< T < T_{CG}$, the average spin overlap $\Ave{q_{\alpha\beta}}$ vanishes as in the high-temperature paramagnetic phase. Meanwhile, the $Z_2$ spin-reflection symmetry is spontaneously broken in the CG phase, which implies that the determinant of the spin-overlap tensor, $\det{q_{\alpha\beta}}$, takes a nonzero value. Denoting this symmetry-breaking bias coming from the CG order by $h(T)$, we may write the distribution of the spin-overlap tensor $q_{\alpha\beta}$ as
\be
P(\{q_{\alpha\beta}\})\propto 
e^{
  -\frac{N}{2\sigma^2}\lb \sum_{\alpha,\beta}q_{\alpha\beta}^2\rb-h(T)\det{q_{\alpha\beta}} 
}.
\ee
Since the standard SG order is absent in the CG phase, the average of any simple moment such as $\Ave{q_{\alpha\beta}^2}$ vanishes in the thermodynamic limit, which is reflected in the first term in the exponent. Using this distribution, the SG Binder parameter is calculated as
\be
g_{SG}=-\frac{1}{4}h^2(T).
\ee
Hence, we find that the SG Binder parameter takes a negative value in the region $T_{SG}< T < T_{CG}$ in spite of the absence of the long-range SG order. Although the detailed form of $h(T)$ is unclear, $h(T)^2$ would grow continuously and  monotonically from zero when the temperature $T$ is decreased across $T=T_{CG}$. It should be noted that, in the CG state of the 3D Heisenberg SG, $g_{SG}$ still remains to be zero in sharp contrast to the {\it XY\/} SG case \cite{Okubo}.

In the SG ordered state realized at $T<T_{SG}$, $g_{SG}$ would take a different form. If there would be no RSB in the SG ordered state, $g_{SG}$ in the thermodynamic limit would jump to unity below $T_{SG}$. If there occurs the 1RSB as in the present model, $g_{SG}$ would take a nontrivial value not equal to unity below $T_{SG}$, eventually approaching to unity in the $T\rightarrow 0$ limit. We show in  \Rfig{gSG-shape} a schematic shape of $g_{SG}$ in the thermodynamic limit  expected in the present model.
\begin{figure}[htbp]
\includegraphics[width=1\columnwidth,height=0.65\columnwidth]{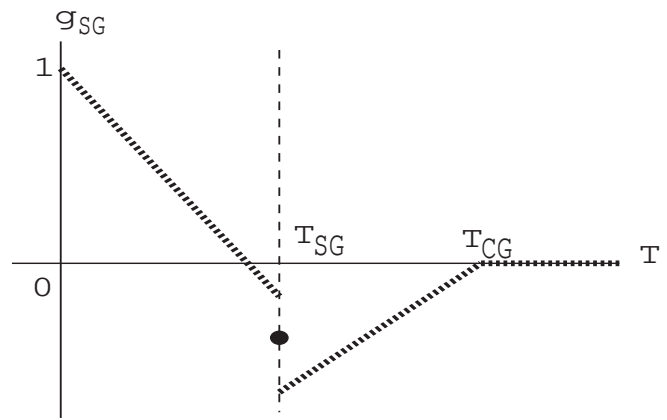}
\caption{\Lfig{gSG-shape} A schematic form of the SG Binder parameter $g_{SG}$ in the thermodynamic limit, as expected in the present model.
}
\end{figure}

 \Rfig{gSG-shape} may enable us to extract further information about the transition temperatures from the data of the SG Binder parameter $g_{SG}$ for larger sizes. We see several characteristic temperatures in $g_{SG}$ for larger sizes, displayed in the left panel of \Rfig{g}. For example, we see in the data of $L=40$ two extrema and an inflection point in between.  In the thermodynamic limit, the extremum at a higher temperature would converge to $T_{CG}$, while the one at a lower temperature to $T_{SG}$. Unfortunately, the extrema are still faint and visible only for larger lattices, making a reliable estimate of  $T_{CG}$ and  $T_{SG}$ difficult. The inflection point would be located somewhere between $T_{SG}$ and $T_{CG}$ in the thermodynamic limit, and thus, can be utilized to give a lower bound of $T_{CG}$ and an upper bound of $T_{SG}$. Indeed, the inflection point of the  $L\geq 20$ data lie around $T\sim 0.31$, which is compatible with our present estimates of $T_{SG}=0.275^{+0.013}_{-0.052}$ and $T_{CG}=0.313^{+0.013}_{-0.018}$.

\end{document}